\documentclass{article}
\usepackage{iclr2027_conference,times}
\iclrfinalcopy

\usepackage{amsmath,amsfonts,bm}

\def\eqref#1{equation~\ref{#1}}

\def\1{\bm{1}}

\DeclareMathAlphabet{\mathsfit}{\encodingdefault}{\sfdefault}{m}{sl}
\SetMathAlphabet{\mathsfit}{bold}{\encodingdefault}{\sfdefault}{bx}{n}

\usepackage{graphicx}
\usepackage{booktabs}
\usepackage{wrapfig}

\usepackage{algorithm}
\usepackage{algpseudocode}

\usepackage{subcaption}

\usepackage{tikz}
\usetikzlibrary{arrows.meta,positioning,calc}

\usepackage{url}
\usepackage{hyperref}
\hypersetup{hidelinks}

\usepackage{tabularx}
\usepackage{array}

\fancypagestyle{labpaperfirst}{
  \fancyhf{}
  \fancyhead[L]{\includegraphics[width=4.2in]{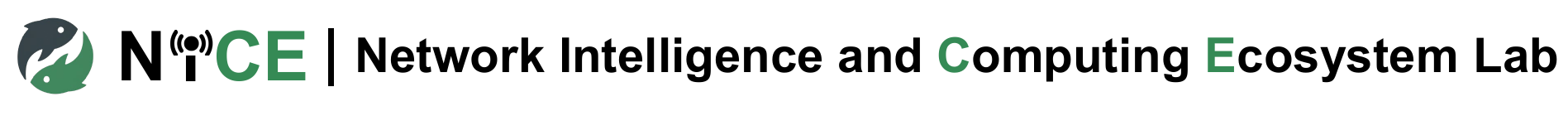}}
  \renewcommand{\headrulewidth}{0pt}
}

\title{NebulaSD: Many-for-Many Speculative Decoding}

\author{
\bfseries Junhao He\textsuperscript{1},\enspace Hongyang Du\textsuperscript{1}\thanks{Corresponding author.}\\[5pt]
{\normalfont\normalsize\textsuperscript{1}The University of Hong Kong}
}

\begin{document}

\vspace*{-20pt}
\maketitle
\thispagestyle{labpaperfirst}
\lhead{Preprint}
\renewcommand{\headrulewidth}{0.4pt}
\vspace{-8pt}

\begin{abstract}
Speculative decoding accelerates Large Language Model (LLM) inference by using a lightweight draft model to propose candidate tokens for parallel verification by a target model. Drafting and verification, however, exhibit different service characteristics and favor different batch configurations, making fixed draft-target coupling inefficient under concurrent workloads. Existing distributed designs can physically separate the two stages, but often retain request or batch affinities that prevent their capacities from being shared globally. We present NebulaSD, a many-for-many, or M-for-N, speculative decoding system that organizes draft and target workers into independently schedulable resource pools and dynamically reconstructs stage-specific batches from shared request pools. Such dynamic reassignment removes fixed worker locality, requiring request states to be made available at newly selected workers without introducing migration stalls. NebulaSD addresses this challenge through worker-triggered batch reconstruction and asynchronous KV-state preparation overlapped with model execution. We evaluate NebulaSD from both system and scaling perspectives, showing that dynamic pooling improves request-round processing rate by $50.4\%$ over a physically disaggregated baseline and $72.6\%$ over co-located execution on a four-GPU deployment while substantially increasing effective GPU utilization. Profile-driven simulations further show approximately proportional compute-side capacity scaling under idealized state movement.
\end{abstract}


\section{Introduction}
Large Language Models (LLMs) power a wide range of interactive applications, but their autoregressive decoding process incurs high generation latency and substantial serving costs~\citep{brown2020language,chowdhery2023palm}. Speculative decoding accelerates generation by using a lightweight draft model to propose multiple candidate tokens and verifying them in parallel with the target model, allowing multiple tokens to be committed within a single target-model forward pass~\citep{leviathan2023fast,chen2023accelerating}.

Existing studies have substantially improved speculative decoding through more accurate candidate generation, efficient verification, and adaptive speculation strategies~\citep{miao2024specinfer,cai2024medusa,li2024eagle}. Recent systems further coordinate speculative execution with dynamic workloads, hardware conditions, and service-level objectives~\citep{liu2023online,huang2025adaspec,li2026adaserve}. Despite these advances, many speculative decoding systems still organize drafting and verification around fixed request or batch associations between the two stages~\citep{tang2026minedraft}. Such coupling can become inefficient when drafting and verification exhibit different service characteristics, because the processing capacity available at one stage cannot be independently matched to the workload of the other.

\begin{figure}[t]
    \centering

    \begin{subfigure}[t]{0.31\linewidth}
        \centering
        \includegraphics[width=\linewidth]{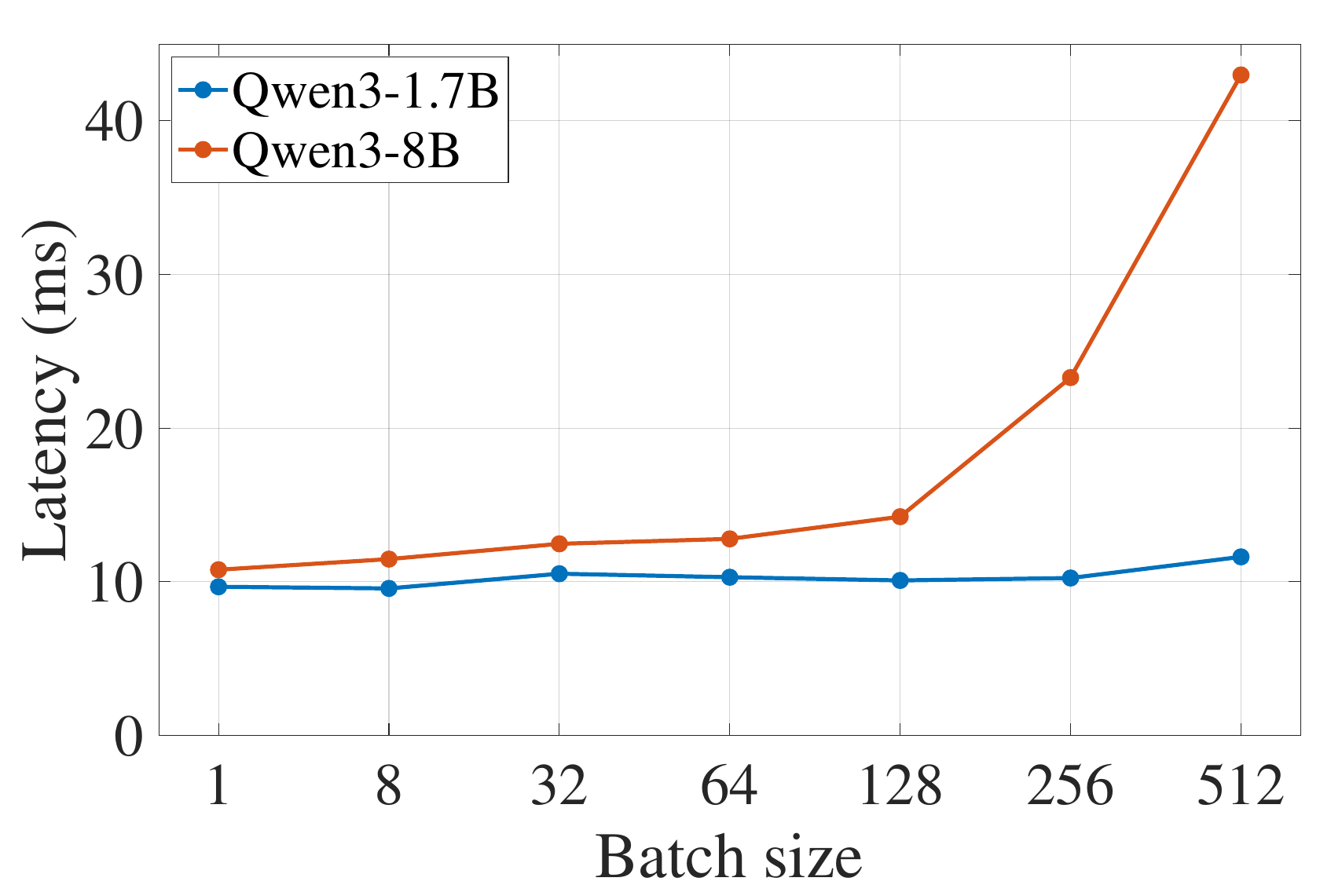}
        \caption{}
        \label{fig:latency}
    \end{subfigure}
    \hfill
    \begin{subfigure}[t]{0.34\linewidth}
        \centering
        \includegraphics[width=\linewidth]{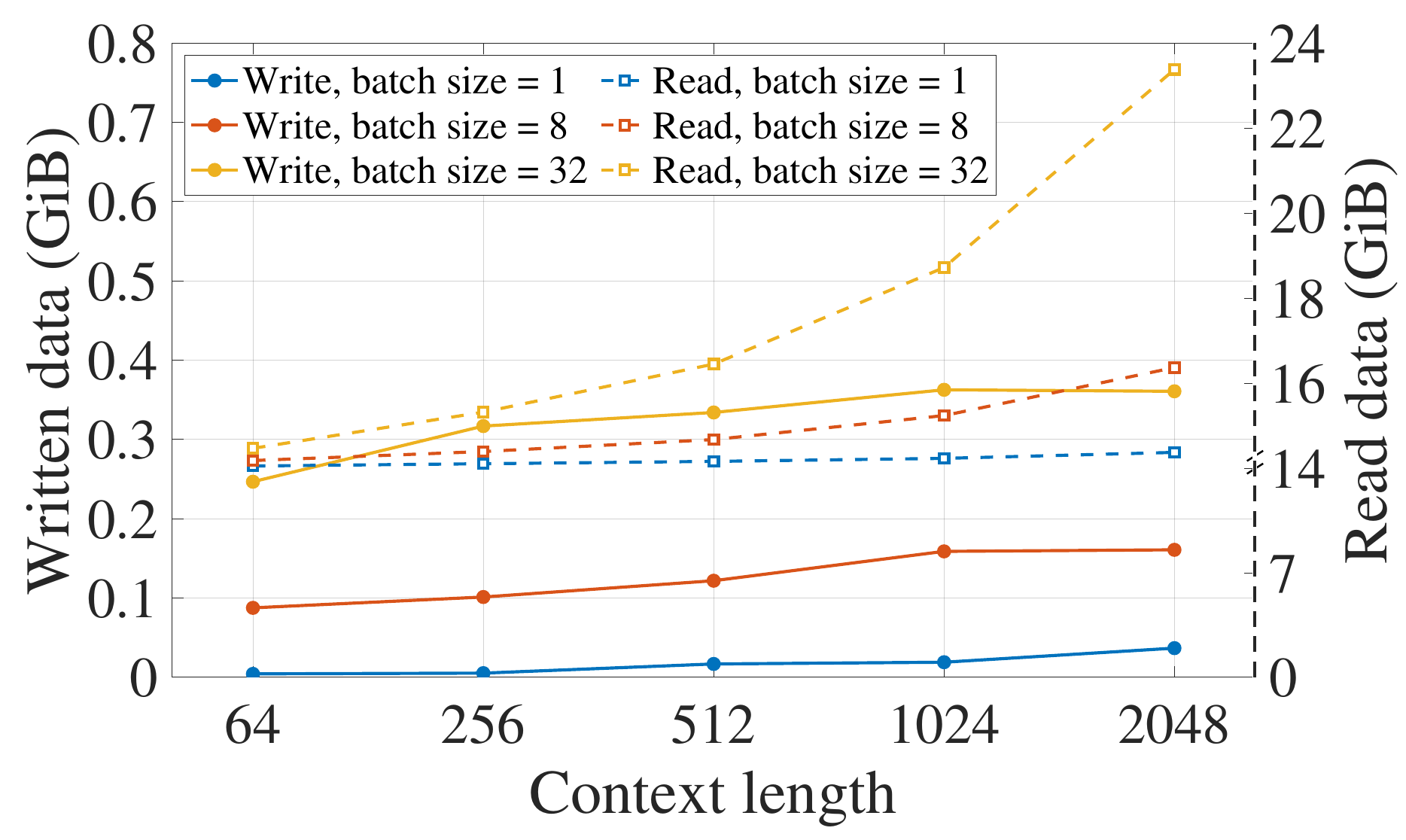}
        \caption{}
        \label{fig:HBM}
    \end{subfigure}
    \hfill
    \begin{subfigure}[t]{0.33\linewidth}
        \centering
        \includegraphics[width=\linewidth]{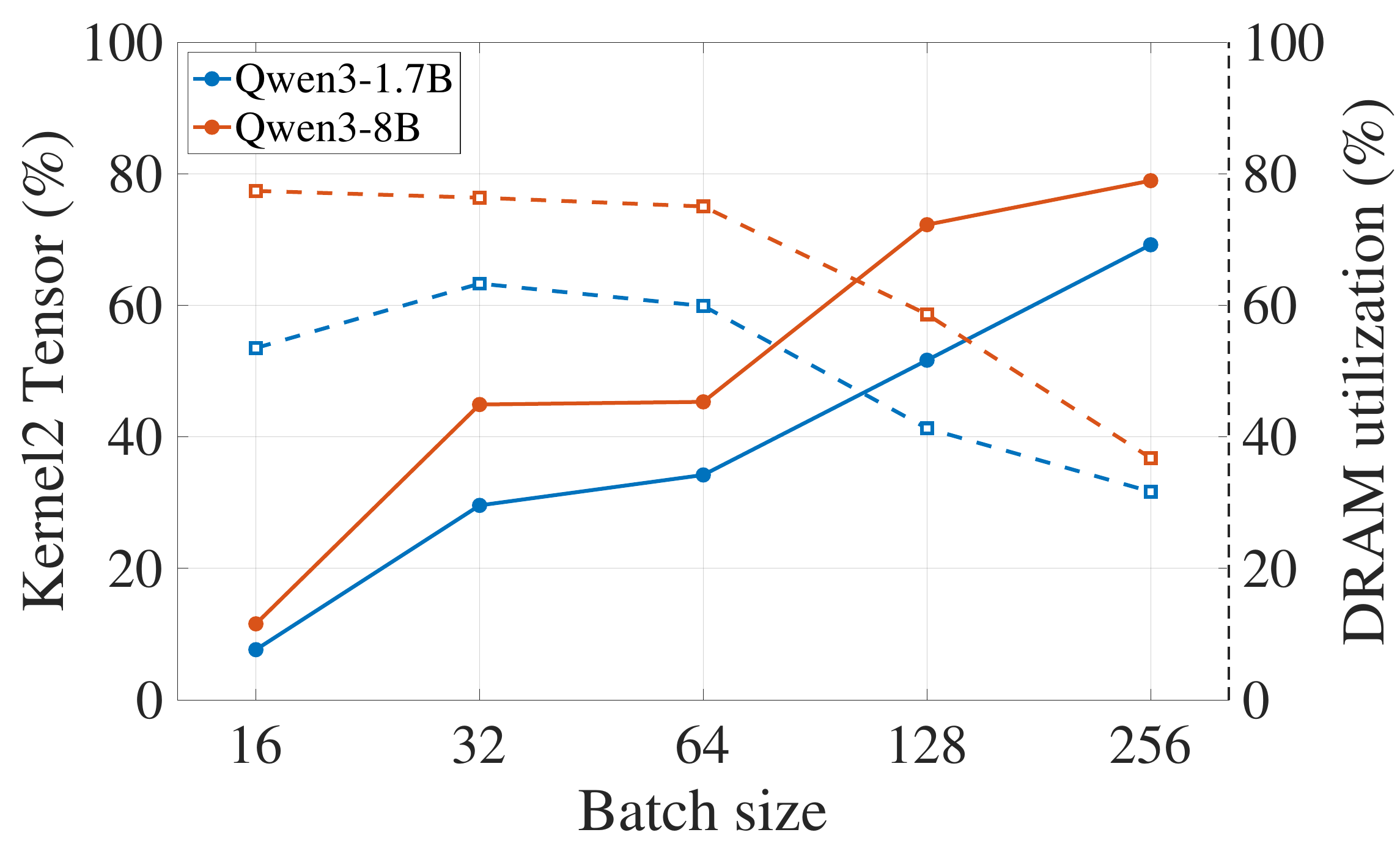}
        \caption{}
        \label{fig:dram_kernel}
    \end{subfigure}

    \vspace{-6pt}
    \caption{
        Forward-pass performance and GPU-memory traffic analysis.
        (a) Forward latency comparison between Qwen3-1.7B and Qwen3-8B
        under different batch sizes.
        (b) GPU-memory read and write traffic of Qwen3-8B under various
        context lengths and batch sizes.
        (c) Kernel2 DRAM\% and Tensor\% that demonstrates memory-bandwidth pressure and tensor core saturation of the dominant compute kernel.
    }
    \label{fig:forward_analysis}
    \vspace{-10pt}
\end{figure}

This difference in service characteristics is evident in how the two stages scale with batch size and utilize GPU resources. Although the draft model is typically much smaller than the target model~\citep{leviathan2023fast,li2024eagle}, drafting and verification follow different execution patterns and scale differently with batch size. As shown in Fig.~\ref{fig:latency}, the forward latency of Qwen3-8B~\citep{yang2025qwen3} increases sharply when the batch size reaches 512, whereas the latency of Qwen3-1.7B~\citep{yang2025qwen3} remains nearly unchanged. Fig.~\ref{fig:dram_kernel} further shows that the two models exhibit different GPU resource demands as the batch size increases. These observations indicate that drafting and verification generally favor different batch configurations and therefore need not provide matched service capacities under the same resource allocation. Under fixed one-to-one pairing, such service-capacity mismatch remains local to each pair. Excess capacity at one stage cannot be shared with requests whose paired counterpart has become the bottleneck, and the effective service rate of each pair is therefore limited by its locally slower stage. As a result, available capacity can remain idle even when other pairs in the system are overloaded.

A natural step toward alleviating this structural inefficiency is to physically separate drafting and verification resources, allowing the two stages to be provisioned according to their different service capacities. Recent distributed speculative decoding systems have explored such separation for heterogeneous accelerator utilization, edge-server collaboration, and shared or centralized verification~\citep{shi2025disaggregated,li2025sled,yu2025dsd,li2026wisp}. Physical disaggregation, however, does not provide global capacity sharing across both stages by itself. Existing designs primarily focus on deployment heterogeneity, remote drafting, verification-side batching, or adaptive speculation instead of exposing both draft and target workers as globally schedulable resource pools in which requests can be reassigned and batches can be reconstructed independently at each stage. Fully exploiting independently provisioned resources therefore requires a stronger serving abstraction that removes fixed worker and cross-stage batch affinities, allowing requests and execution capacity to be shared dynamically within each stage according to workload and resource availability.

However, such dynamic request sharing requires requests to be reassigned across workers, which in turn breaks the state locality provided by fixed worker assignment. During auto-regressive generation, each request carries model state in the form of a growing KV cache~\citep{kwon2023efficient,lee2024infinigen}. Once requests can be reassigned across draft or target workers, the corresponding state must also be made available at the selected destination before execution can proceed. Performing state movement only after worker assignment places migration directly on the execution critical path, whereas preparing it too late can leave otherwise available workers idle. At the same time, aggressively moving state in advance may introduce unnecessary transfers and interfere with foreground computation. Efficient M-for-N execution therefore requires worker assignment, batch formation, and state preparation to be coordinated so that the required state becomes available when execution is ready without turning state migration into a new system bottleneck.

To address these coupled requirements, we present NebulaSD, an M-for-N speculative decoding system that organizes $M$ draft workers and $N$ target workers as two independently provisioned and globally schedulable resource pools. Instead of permanently binding requests to individual draft-target pairs or preserving cross-stage batch affinity, NebulaSD dynamically reconstructs worker-local batches from shared request pools, allowing draft and target capacity to be independently matched to their workloads. To support such dynamic reassignment, NebulaSD coordinates worker-triggered batch reconstruction with advance state preparation, overlapping KV-state movement with foreground model execution whenever possible. Together, these mechanisms enable flexible resource sharing across both worker pools while keeping most state-movement overhead off the execution critical path. Our contributions are summarized below.

\begin{itemize}

\item We propose an M-for-N speculative decoding architecture that decouples draft and target workers into independently schedulable resource pools, removing fixed request-worker and cross-stage batch affinities and enabling capacity sharing across the two stages.

\item We develop a worker-triggered stage-wise batch reconstruction policy that reconstructs each worker's upcoming batch from a shared request pool according to execution readiness, state availability, and service efficiency, enabling fine-grained dynamic assignment without global repartitioning.

\item We design an asynchronous KV-state preparation mechanism that overlaps state movement with model computation through advance prefetching, incremental writeback, and double buffering, reducing the runtime overhead introduced by dynamic request reassignment.

\end{itemize}

\section{Related Work}


\paragraph{Distributed speculative decoding.} Speculative decoding accelerates autoregressive generation by using a lightweight draft model to propose candidate tokens and a target model to verify them in parallel~\citep{leviathan2023fast,chen2023accelerating}. Recent work has begun to separate draft generation and target verification across devices or serving instances. Disaggregated speculative decoding places draft and target models on heterogeneous GPUs for improved hardware reuse and carbon efficiency~\citep{shi2025disaggregated}. SLED and WISP study edge-side drafting with shared server-side verification, including cross-request batching and SLO-aware scheduling~\citep{li2025sled,li2026wisp}. DSD considers edge-cloud execution under network and batching effects with adaptive speculation control~\citep{yu2025dsd}, while SPECTRE exploits underutilized remote model services as drafters in multi-model cloud serving~\citep{xie2026spectre}. These systems primarily target heterogeneous deployment, edge-cloud collaboration, verification scheduling, or remote drafting, rather than globally pooling and independently scheduling both draft and target resources.

\paragraph{LLM serving.} LLM serving systems improve inference efficiency through batching, memory management, scheduling, and SLO-aware resource allocation. Orca introduces iteration-level scheduling for generative inference~\citep{yu2022orca}, while vLLM improves KV-cache management with PagedAttention to reduce memory fragmentation and support larger batches~\citep{kwon2023efficient}. Sarathi-Serve further improves the throughput-latency trade-off through chunked prefill and stall-free scheduling~\citep{agrawal2024taming}. More recent systems explicitly consider latency objectives and workload heterogeneity: DistServe disaggregates prefill and decode under TTFT and TPOT constraints~\citep{zhong2024distserve}, Llumnix dynamically reschedules requests across model instances~\citep{sun2024llumnix}, and JITServe studies SLO-aware scheduling with imprecise request information~\citep{zhang2026jitserve}. These systems improve resource utilization and request scheduling in general LLM serving, but do not directly address the asymmetric service characteristics and alternating execution dependencies between drafting and verification in speculative decoding.

\paragraph{KV cache migration.} KV cache movement is a key challenge for dynamic LLM serving because each active request carries state that grows with sequence length. Llumnix supports live migration of active requests across model instances~\citep{sun2024llumnix}, while SpotServe introduces state migration and recovery for preemptible GPU serving~\citep{miao2024spotserve}. Prefill-decode disaggregated systems such as Splitwise and DistServe transfer request states across stages, making performance sensitive to communication bandwidth and placement decisions~\citep{patel2024splitwise,zhong2024distserve}. Other systems reduce KV-transfer or restoration overhead through compression and streaming~\citep{liu2024cachegen}, distributed KV-cache storage~\citep{qin2025mooncake}, or selective recomputation with cached-KV loading~\citep{yao2025cacheblend}. These mechanisms reduce state-movement cost, but do not directly coordinate frequent cross-worker state preparation with stage-specific batch reconstruction in speculative execution.


\section{Preliminary}

\paragraph{GPU-Memory characteristics of LLM inference.}
LLM decoding at small-to-moderate batch sizes often exhibits low arithmetic intensity and is sensitive to GPU-memory bandwidth~\citep{agrawal2024taming}. During each target forward, model weights and the KV states associated with the existing context are repeatedly accessed, while new KV entries are generated only for the tokens processed in the current iteration. As shown in Fig.~\ref{fig:HBM}, GPU-memory traffic is highly asymmetric. The amount of data read is substantially larger than the amount written across all evaluated context lengths and batch sizes. 

\paragraph{KV cache organization.}
Modern LLM serving systems commonly use paged KV-cache management to reduce memory fragmentation and support dynamic batching~\citep{kwon2023efficient}. However, a request's KV state may span multiple non-contiguous blocks, making migration require block-map traversal, destination allocation, metadata reconstruction, and fragmented copies. KV cache is also append-dominant: across consecutive decoding or verification iterations, most prefix states remain unchanged while only newly generated states are appended. These properties make KV-cache movement more complex than transferring a single contiguous tensor.

\section{System Model}
\label{sec:system_model}
\begin{figure}[t]\vspace{-10pt}\centering
    \includegraphics[width=0.9\linewidth]{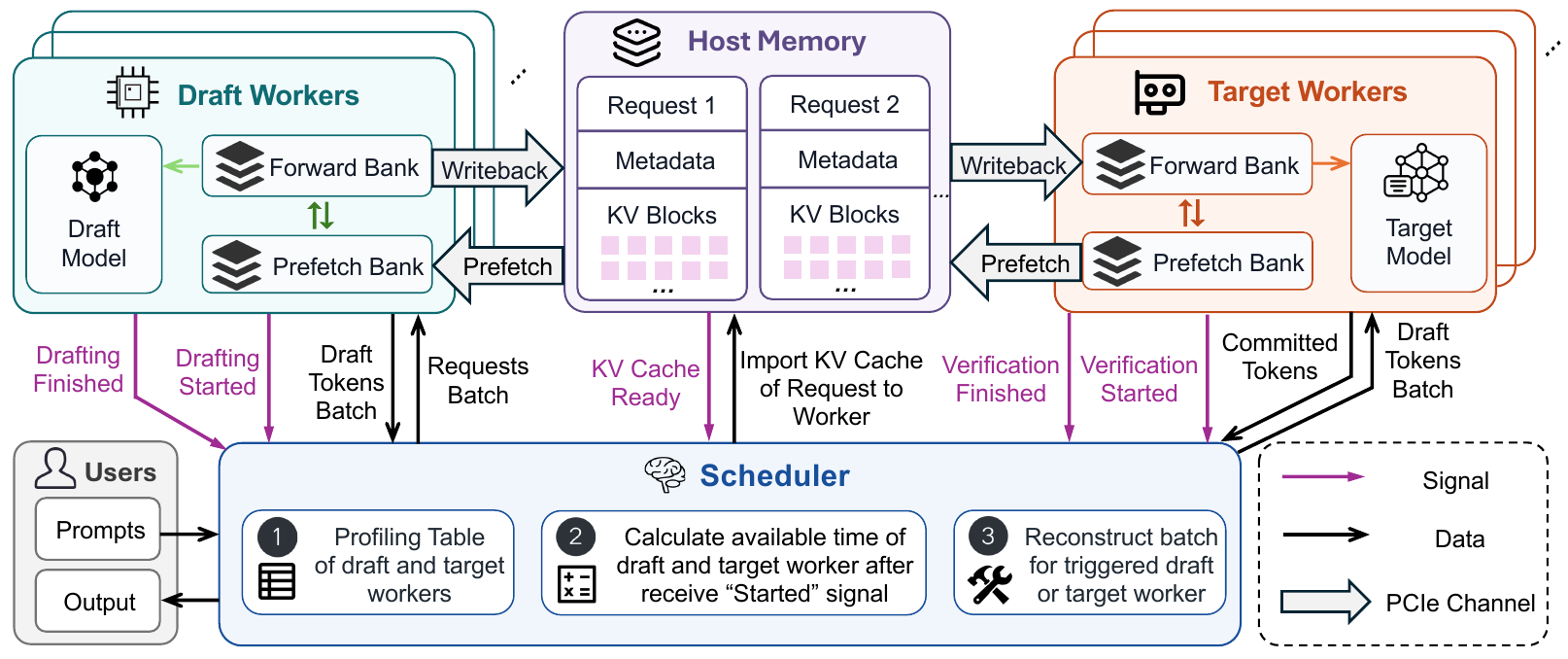}
    \caption{Architecture of NebulaSD. Draft and target workers form independent resource pools, while a centralized scheduler reconstructs worker-local batches from shared request pools and uses the host-side KV store to support cross-worker state preparation.}    
    \label{fig:architecture}
    \vspace{-10pt}
\end{figure}

We first characterize the stage-level service capacity enabled by pooled execution, then describe how NebulaSD plans future execution and reconstructs worker-local batches under request and state readiness constraints. As illustrated in Fig.~\ref{fig:architecture}, NebulaSD consists of a shared draft-worker pool $\mathcal{D}=\{d_1,\ldots,d_M\}$, a shared target-worker pool $\mathcal{T}=\{t_1,\ldots,t_N\}$, a centralized scheduler, and a host-side KV store. We use $s\in\{\mathrm{D},\mathrm{T}\}$ to denote an execution stage and $\mathcal{W}_s$ to denote its corresponding worker pool, where $\mathcal{W}_{\mathrm{D}}=\mathcal{D}$ and $\mathcal{W}_{\mathrm{T}}=\mathcal{T}$.

Each request $r$ maintains two logical execution states, $S_{\mathrm{D}}^{r}$ and $S_{\mathrm{T}}^{r}$, corresponding to the draft and target models, respectively. Requests are not permanently bound to individual workers. Instead, each stage maintains a shared request pool, from which different workers may serve a request across successive speculative iterations. The required model state is prepared through the host-side KV store before execution, allowing request execution to move across workers without preserving fixed worker affinity. 

NebulaSD also removes cross-stage batch affinity. Requests processed together in one stage are not required to remain together in the next stage; workers instead reconstruct their upcoming batches independently from the corresponding stage-level pool. This worker-local reconstruction decouples batch formation across stages while preserving global request sharing within each worker pool. The scheduling policy for constructing such batches is defined below.

\paragraph{Stage progression and pool capacity.}
Each request alternates between the draft and target stages across speculative iterations. For request $r$ in iteration $q$, the execution dependency is
\begin{equation}
\mathrm{draft}_{r}^{q}
\prec
\mathrm{target}_{r}^{q}
\prec
\mathrm{draft}_{r}^{q+1},
\label{eq:stage_dependency}
\end{equation}
where $\prec$ denotes an execution dependency. The output of the predecessor stage must be available before the next-stage computation can begin, although the corresponding worker assignment and KV-state preparation may be planned earlier. Let $\widehat T_s(\mathcal{B})=\phi_s(\mathbf{z}_{\mathcal{B}})$ denote the profiled execution time of batch $\mathcal{B}$ at stage $s$, where $\mathbf{z}_{\mathcal{B}}$ summarizes the workload characteristics relevant to execution, such as batch size, context length, and proposal depth. The corresponding stage service rate is approximately
\begin{equation}
\mu_s(\mathcal{B})
=
\frac{|\mathcal{B}|}{\widehat T_s(\mathcal{B})}.
\label{eq:worker_service_rate}
\end{equation}
Thus, the aggregate service capacity of a worker pool depends on both its number of workers and the batch configurations that those workers can efficiently execute. Since every speculative iteration requires both draft and target execution, the steady-state processing rate is ultimately limited by the slower stage,
\begin{equation}
\mu_{\mathrm{sys}}
\leq
\min\{\mu_{\mathrm{D}},\mu_{\mathrm{T}}\}.
\label{eq:system_capacity}
\end{equation}
This motivates independently provisioning and scheduling the two pools according to their different service capacities instead of preserving fixed draft-target pairs.

\begin{figure}[t]
\vspace{-10pt}
\centering
    \includegraphics[width=0.9\linewidth]{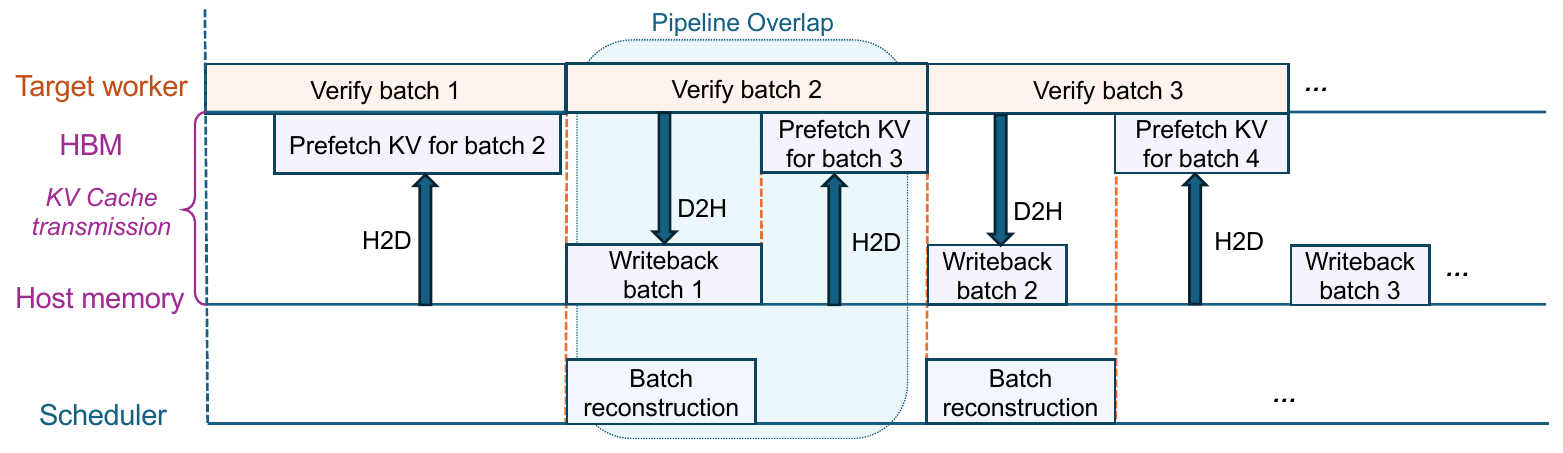}
    \caption{Pipeline of a target worker in NebulaSD.}
    \label{fig:pipeline}
\vspace{-10pt}
\end{figure}

\paragraph{Overlapped batch planning and state preparation.}
Given the stage dependency in Eq.~\ref{eq:stage_dependency}, NebulaSD applies the same scheduling and preparation abstraction to both draft and target execution. In particular, either a draft or target worker can independently trigger planning for its next batch, select requests from the corresponding shared request pool, and initiate state preparation before the predecessor stage completes. Fig.~\ref{fig:pipeline} illustrates this symmetric pipeline from the perspective of a target worker. Although the next-stage computation cannot begin until its predecessor output becomes available, NebulaSD does not require worker assignment and state preparation to wait for that completion. When a worker obtains an opportunity to prepare its next batch, the scheduler may use predicted predecessor completion times to select requests and prepare the required state in advance.

Consequently, target computation can overlap with draft-state preparation, while draft computation can overlap with target-state preparation. The steady-state execution therefore alternates between $\textsc{target}^{q}$, $\textsc{draft}^{q+1}$, and $\textsc{target}^{q+1}$, while the state preparation for each upcoming stage is overlapped with the computation of its predecessor. A prepared batch is launched only when its predecessor output, required KV state, and destination worker are all actually ready. Predicted completion times are therefore used for advance planning and preparation, whereas runtime readiness events determine when execution can safely begin.

\paragraph{Worker-triggered batch reconstruction.} Consider a scheduling opportunity for worker $w\in\mathcal{W}_s$ at stage $s\in\{\mathrm{D},\mathrm{T}\}$, and let $\bar{s}$ denote its predecessor stage, i.e., $\bar{s}=\mathrm{T}$ for $s=\mathrm{D}$ and $\bar{s}=\mathrm{D}$ for $s=\mathrm{T}$. Since the destination worker is fixed by the scheduling opportunity, the scheduler reconstructs its next batch from the eligible requests $\mathcal{E}\subseteq\mathcal{R}_s$. NebulaSD may plan this batch before all predecessor computations finish and therefore distinguishes observed completion times from predicted ones. If the predecessor of request $r$ has completed, its observed completion time $e_r$ is used directly. Otherwise, for a predecessor batch that started at $t_{\mathrm{start}}$ with workload descriptor $\mathbf{z}$, the scheduler predicts
\begin{equation}
\widehat e_r=t_{\mathrm{start}}+\phi_{\bar{s}}(\mathbf{z}),
\label{eq:completion_prediction}
\end{equation}
using the profiling table of the predecessor stage. Let $\widetilde e_r$ denote the predecessor completion time available to the scheduler, using $e_r$ when observed and $\widehat e_r$ otherwise. The predicted availability $\widehat a_w$ of the destination worker is analogously obtained from the start time and profiled duration of its currently executing stage-$s$ batch. For state readiness, let $\widehat k_r(w)$ denote the predicted time at which the KV state required by request $r$ becomes available on worker $w$, based on its current state location and the estimated preparation or migration cost. The state-ready time of a candidate batch is therefore
\begin{equation}
\widehat k(\mathcal{B},w)=\max_{r\in\mathcal{B}}\widehat k_r(w).
\label{eq:batch_kv_ready}
\end{equation}
The earliest predicted start time of $\mathcal{B}$ is
\begin{equation}
\widehat S(\mathcal{B},w)=\max\left\{\widehat a_w,\,\max_{r\in\mathcal{B}}\widetilde e_r,\,\widehat k(\mathcal{B},w)\right\}.
\label{eq:batch_start}
\end{equation}
This formulation captures the three conditions that jointly determine future execution: worker availability, predecessor-output readiness, and KV-state readiness.

\paragraph{Service-constrained batch reconstruction.}
Batch reconstruction improves stage efficiency by aggregating requests into favorable batch configurations, but exhaustive subset optimization at every worker opportunity would introduce scheduling overhead on the serving critical path. NebulaSD therefore adopts a bounded online reconstruction policy that prioritizes timely service while preserving batching efficiency. Let $S_r$ denote the start time of request $r$ at its next stage and $e_r$ the completion time of its predecessor stage, giving the realized service interval $G_r=S_r-e_r$. For each stage $s$, NebulaSD specifies a configurable target service interval $g_s$, which determines the nominal timescale at which requests should receive their next-stage service. At scheduling time, the predecessor completion time is represented by $\widetilde e_r$, while $\widehat S(\{r\},w)$ denotes the predicted start time if $r$ were served alone by the triggered worker $w$. We define the corresponding service-admission bound as
\begin{equation}
\widehat d_r(w)=\max\left\{\widetilde e_r+g_s,\widehat S(\{r\},w)\right\}+\Delta_s,
\label{eq:service_admission_bound}
\end{equation}
where the configurable slack $\Delta_s$ controls the additional delay permitted for batch formation. A reconstructed batch $\mathcal{B}$ is admissible on worker $w$ only if
\begin{equation}
|\mathcal{B}|\leq B_{\max},\qquad \mathrm{Fit}(\mathcal{B},w)=1,\qquad \widehat S(\mathcal{B},w)\leq\min_{r\in\mathcal{B}}\widehat d_r(w).
\label{eq:batch_reconstruction}
\end{equation}
The first two conditions enforce batch-size and physical-capacity constraints, while the last bounds batching-induced delay for every included request. NebulaSD then ranks eligible requests by service urgency and temporal compatibility and greedily constructs a batch subject to these constraints, avoiding the cost of exhaustive subset search in the online scheduling path. The waiting-time component of the ranking provides an aging effect for deferred requests, while the predicted singleton start time $\widehat S(\{r\},w)$ provides a feasibility baseline, preventing unavoidable worker, predecessor, or state-readiness delays from being counted as batching-induced delay.

\begin{algorithm}[t]
\caption{Worker-triggered batch reconstruction}
\label{alg:stage_scheduling}
\footnotesize
\begin{algorithmic}[1]
\Require Stage $s$, triggered worker $w$, request pool $\mathcal{R}_s$
\State Filter eligible requests and predict request/worker readiness
\State Rank candidates by predecessor readiness, waiting time, and temporal compatibility
\State $\mathcal{F}\gets$ first at most $B_{\max}$ candidates in the ranked order; $\mathcal{B}\gets\emptyset$
\ForAll{$r\in\mathcal{F}$ in the ranked order}
    \If{$\mathcal{B}\cup\{r\}$ satisfies Eq.~\ref{eq:batch_reconstruction}}
        \State $\mathcal{B}\gets\mathcal{B}\cup\{r\}$
    \EndIf
\EndFor
\State Freeze/reserve nonempty $\mathcal{B}$ and issue \textsc{Prepare}$(\mathcal{B},w)$
\State \textsc{Run} when $\textsc{InputReady}\land\textsc{KVReady}\land\textsc{WorkerReady}$
\end{algorithmic}
\end{algorithm}

\section{Implementation}
\label{sec:implementation}

\paragraph{Runtime architecture.} NebulaSD implements target workers on top of SwiftLLM~\citep{liu2024swiftllm} and exposes a unified draft-worker interface that can accommodate different speculative-decoding backends. Both worker pools provide the same scheduler-facing abstractions for batch execution, KV-state migration, and asynchronous event reporting. A centralized scheduler maintains request, worker, and state-residency metadata, receives execution-start and completion signals from both pools, and triggers worker-local batch reconstruction and state preparation according to the scheduling opportunities described in Section~\ref{sec:system_model}. This design keeps scheduling independent of backend-specific execution while allowing both draft and target requests to migrate across workers between speculative iterations.

\paragraph{Persistent HostKV.}
NebulaSD maintains host-resident KV states for both the draft and target models throughout the lifetime of each active request. The two model states are stored in separate stage-specific regions allocated from pre-registered pinned-memory arenas and identified by stable offsets. Each state is associated with lightweight metadata recording its valid sequence length, storage location, and version. The persistent layout avoids repeated host-buffer allocation, memory registration, and repacking during migration, while allowing the scheduler to prepare either draft-side or target-side state on any worker in the corresponding pool.

\paragraph{Incremental KV writeback and migration estimation.} NebulaSD updates HostKV incrementally by asynchronously writing back only the valid KV suffix that is not already persistent in host memory, while leaving the unchanged prefix intact. For a batch $B$ at stage $s\in\{\mathrm{D},\mathrm{T}\}$, let $\Delta Q_s(B)$ denote the number of KV tokens that must be written back to HostKV, and let $Q_s(B)$ denote the number of KV tokens that must be restored from HostKV during preparation. With stage-specific KV size $b_{\mathrm{KV},s}$ and profiled effective transfer bandwidths, NebulaSD estimates the corresponding migration costs as
\begin{equation}
\widehat{T}_{s,\mathrm{wb}}(B)=\frac{\Delta Q_s(B)b_{\mathrm{KV},s}}{\beta_{s,\mathrm{D2H}}},\qquad \widehat{T}_{s,\mathrm{in}}(B)=\frac{Q_s(B)b_{\mathrm{KV},s}}{\beta_{s,\mathrm{H2D}}}.
\label{eq:kv_migration_cost}
\end{equation}
These estimates, together with HostKV readiness and transfer progress, are used to derive the request-level predicted KV-ready times $\widehat{k}_r(w)$ and hence the batch-level readiness $\widehat{k}(B,w)$ in Eq.~\ref{eq:batch_kv_ready}. Incremental writeback reduces redundant GPU-memory reads and D2H traffic, while asynchronous transfer allows state synchronization to proceed independently of foreground execution.

\paragraph{Overlapped state preparation.}
Each draft and target worker maintains two KV banks so that state preparation for a future batch can overlap with foreground model execution. While one bank serves the current computation, the other can complete state writeback and prepare the KV state required by the next scheduled batch. The banks alternate roles across executions, allowing most state movement to proceed outside the foreground critical path. Detailed bank states, reuse dependencies, and physical-capacity constraints are described in Appendix~\ref{app:double_bank}.

\section{Experiment}
\label{sec:evaluation}
We evaluate NebulaSD from both system and scaling perspectives. We first use a four-GPU deployment to examine whether physical disaggregation alone is sufficient and how dynamic pooling affects system processing capacity, hardware utilization, and per-request service under a fixed resource budget. We then use large-scale profile-driven simulation to study how NebulaSD scales with larger worker pools and how resource provisioning and workload concurrency jointly determine the system's operating point across different deployment scales.

For the four-GPU study, experiments are conducted on four RTX 4090 GPUs using Qwen3-0.6B~\citep{yang2025qwen3} as the draft model and Qwen3-8B~\citep{yang2025qwen3} as the target model, with a speculative depth of $4$. We initialize 512 requests with a maximum generation length of 128 tokens, let them arrive simultaneously, and fix the maximum target batch size to $32$ across all configurations. \emph{Co-located} places one draft instance ($\mathrm{D}$) and one target instance ($\mathrm{T}$) on each GPU and processes requests with the conventional coupled execution model. \emph{Native Distributed} physically separates the two stages into a $2\mathrm{D}+2\mathrm{T}$ deployment with a draft batch-size limit of $128$, but preserves cross-stage batch affinity. NebulaSD uses the same $2\mathrm{D}+2\mathrm{T}$ resource configuration and batch-size limits, while additionally enabling shared request pools, worker-triggered batch reconstruction, cross-worker state migration, and early state preparation. We report the request-round processing rate, defined as the number of completed speculative rounds per second, where each round consists of one draft stage followed by target verification.

\subsection{From Disaggregation to Dynamic Pooling}
\label{sec:eval_four_gpu}
\begin{figure}[!t]\vspace{0pt}\centering
    \begin{subfigure}[t]{0.32\linewidth}
        \centering
        \includegraphics[width=\linewidth]{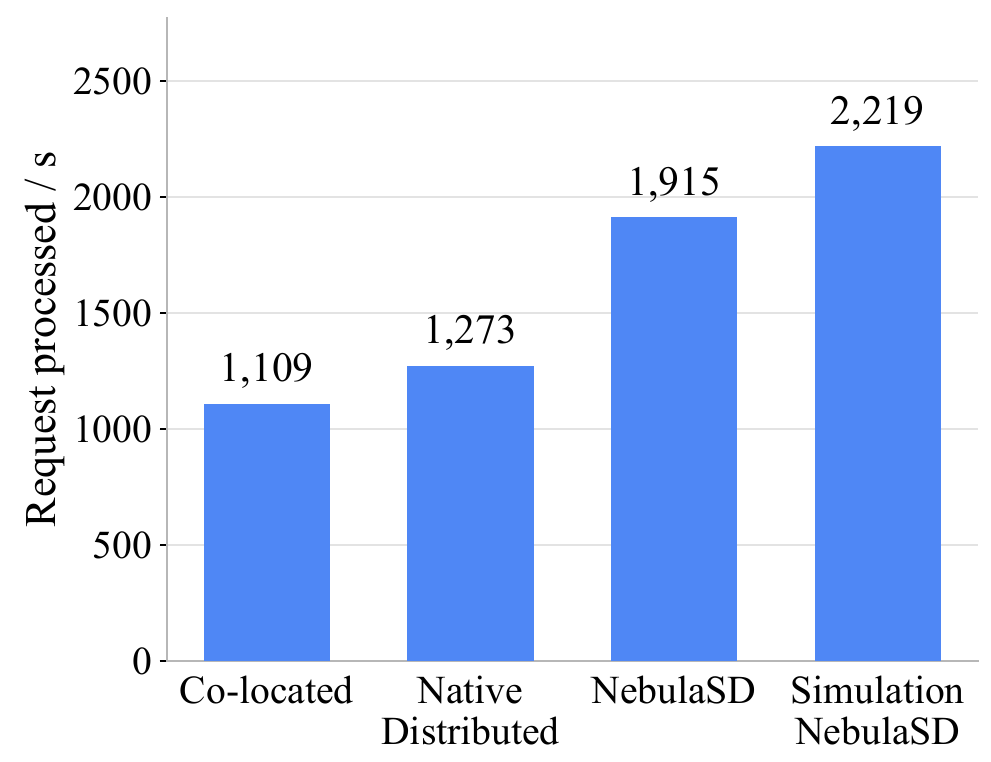}
        {\small (a) Request-round rate}
        \label{fig:sys_request_rate}
    \end{subfigure}
    \hfill
    \begin{subfigure}[t]{0.33\linewidth}
        \centering
        \includegraphics[width=\linewidth]{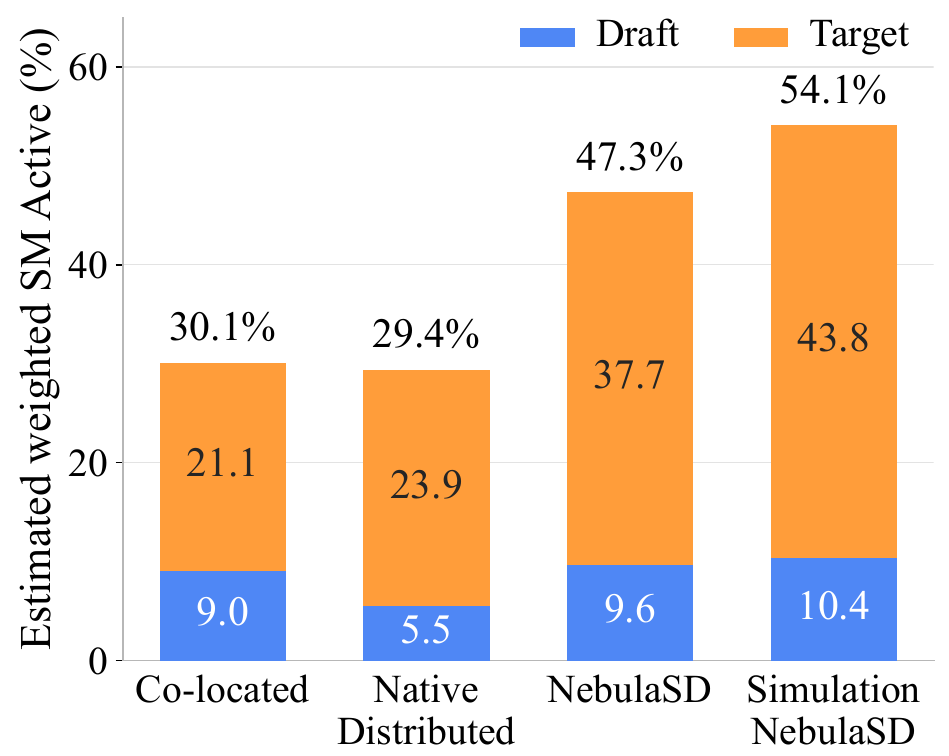}
        {\small (b) Time-weighted SM activity}
        \label{fig:weighted_sm}
    \end{subfigure}
    \hfill
    \begin{subfigure}[t]{0.33\linewidth}
        \centering
        \includegraphics[width=\linewidth]{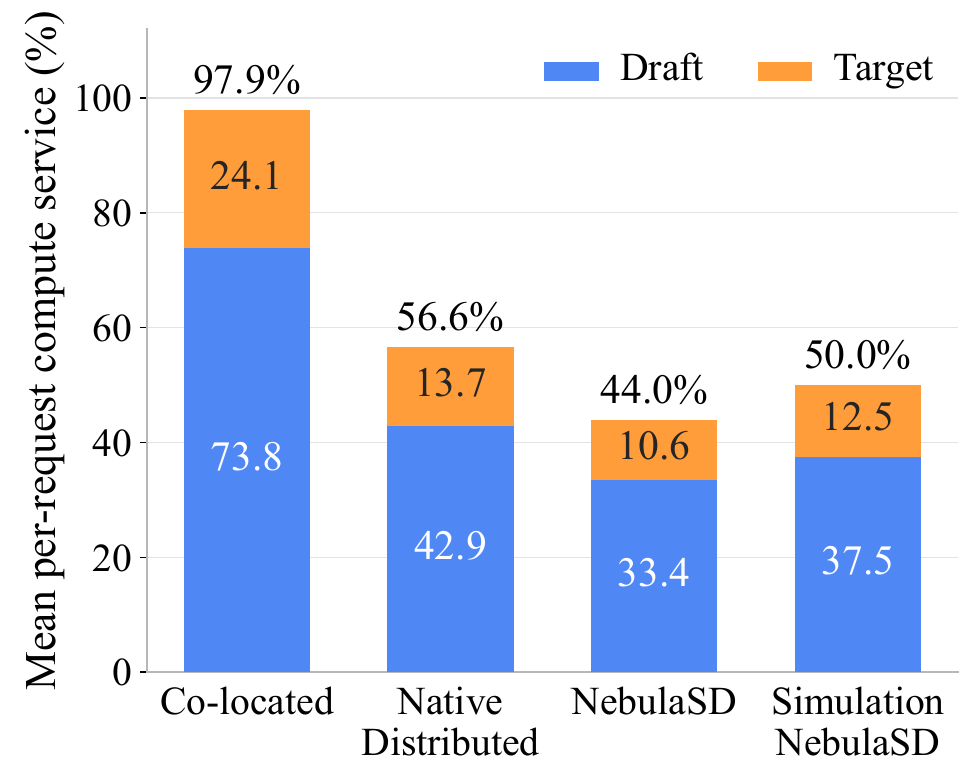}
        {\small (c) Compute-service fraction}
        \label{fig:user_perspective}
    \end{subfigure}

    \vspace{1pt}
    \caption{
    Steady-state comparison of Co-located, Native Distributed, NebulaSD, and simulated NebulaSD. (a) Request-round processing rate, measured in completed speculative rounds per second. (b) Estimated GPU utilization based on compute-time-weighted Streaming Multiprocessor (SM) activity. (c) Mean fraction of time each request receives worker computation.
    }
    \label{fig:real_experiment}
    \vspace{-5pt}
\end{figure}

Fig.~\ref{fig:real_experiment} shows that NebulaSD improves effective hardware utilization and system processing capacity beyond what is achieved by physical disaggregation alone. Compared with both Co-located and Native Distributed execution, NebulaSD keeps the independently provisioned draft and target resources busy for a substantially larger fraction of time, increasing time-weighted SM activity from $30.1\%$ and $29.4\%$ to $47.3\%$, while increasing the request-round processing rate from $1109.37$ and $1273.35$ rounds/s to $1914.83$ rounds/s, respectively. 

\begin{wraptable}{r}{0.62\linewidth}
\vspace{-8pt}
\centering
\caption{Steady-state execution characteristics on four RTX 4090 GPUs.}
\label{tab:steady_four_gpu}
\setlength{\tabcolsep}{3pt}
\renewcommand{\arraystretch}{1.08}
\scriptsize
\resizebox{\linewidth}{!}{%
\begin{tabular}{l|cccc}
\toprule
\textbf{Metric}
& \textbf{Co-located}
& \textbf{Native Dist.}
& \textbf{NebulaSD}
& \textbf{Ideal Sim.} \\
\midrule

\multicolumn{5}{l}{\textit{\textbf{System performance}}} \\
Request rounds (/s)
& 1109.37
& 1273.35
& 1914.83
& 2218.67 \\
\cmidrule(lr){1-5}

\multicolumn{5}{l}{\textit{\textbf{Batching and model execution}}} \\
Draft avg.\ batch
& 32.00
& 128.00
& 89.24
& 96.00 \\
Target avg.\ batch
& 32.00
& 32.00
& 31.32
& 32.00 \\
Drafting time (ms)
& 86.37
& 87.21
& 87.72
& 86.38 \\
Verification time (ms)
& 27.83
& 27.51
& 28.15
& 28.93 \\
\cmidrule(lr){1-5}

\multicolumn{5}{l}{\textit{\textbf{Pipeline utilization}}} \\
Draft compute rate (\%)
& 73.83
& 42.90
& 95.28
& 100.00 \\
Target compute rate (\%)
& 24.11
& 54.67
& 86.34
& 100.00 \\
Draft mean gap (ms)
& 30.20
& 114.96
& 4.38
& 0.00 \\
Target mean gap (ms)
& 89.17
& 24.03
& 4.45
& 0.00 \\
\cmidrule(lr){1-5}

\multicolumn{5}{l}{\textit{\textbf{SM activity during computation}}} \\
Draft SM Active (\%)
& 12.22
& 25.49
& 20.25
& 20.72 \\
Target SM Active (\%)
& 87.51
& 87.51
& 87.32
& 87.51 \\
\bottomrule
\end{tabular}%
}
\vspace{-8pt}
\end{wraptable}

The definition of time-weighted SM activity is given in Appendix~\ref{app:sm_activity}. This improvement comes from dynamically reconstructing batches and filling execution bubbles rather than simply increasing batch size. In fact, NebulaSD uses a smaller average draft batch than Native Distributed, while the target batch size and per-batch computation time remain similar across the two configurations. Instead, the mean gaps between consecutive computations decrease from $114.96$ to $4.38$ ms for draft workers and from $24.03$ to $4.45$ ms for target workers, indicating that dynamic pooling primarily improves performance by keeping both worker pools continuously supplied with executable work. The trade-off is also visible from the request perspective: under the same four-GPU budget, the mean compute-service fraction decreases from $97.9\%$ for Co-located and $56.6\%$ for Native Distributed to $44.0\%$ with NebulaSD, as each request receives a smaller fraction of GPU compute service while the system serves more requests concurrently. In larger deployments, however, this trade-off can be mitigated by scaling the worker pools and better matching the aggregate service capacities of the draft and target stages, which will be further illustrated in the large-scale simulation experiments in Section~\ref{sec:eval_scaling}.

The idealized simulation further isolates the scheduling potential of NebulaSD from practical runtime effects. As shown in Table~\ref{tab:steady_four_gpu}, the real system and simulation exhibit similar per-batch computation costs, while their main difference lies in the idle gaps between consecutive worker computations. In NebulaSD, the execution of a prepared batch is jointly constrained by predecessor-output readiness, worker availability, and KV-state readiness, as modeled in Eq.~\ref{eq:batch_start}. At each worker-triggered scheduling opportunity, the scheduler reconstructs the worker's next batch using the service-aware ranking and admissibility constraints in Eq.~\ref{eq:batch_reconstruction}. Under the idealized simulator, these scheduling opportunities can be exploited with little runtime disturbance, allowing workers to approach continuous execution. The real implementation, in contrast, is affected by KV preparation, asynchronous event timing, scheduling overhead, and prediction mismatch, which introduce additional gaps between computations. Therefore, the difference between real execution and simulation mainly reflects practical orchestration overhead.

\subsection{Simulation for Scaling and Capacity Matching}
\label{sec:eval_scaling}

\begin{figure}[t]\vspace{-10pt}\centering
    \begin{subfigure}[t]{0.33\linewidth}
        \centering
        \includegraphics[width=\linewidth]{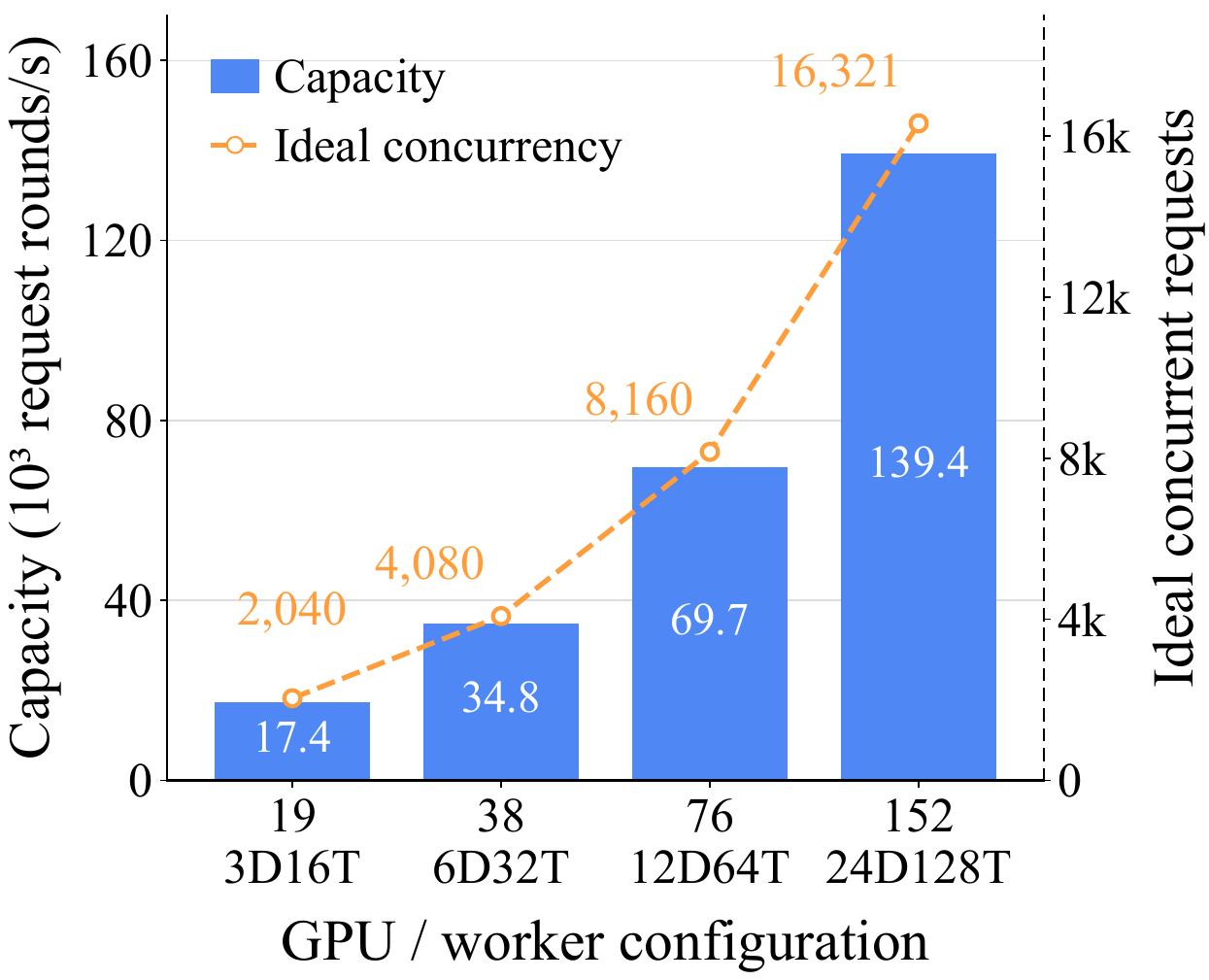}
        \caption{Capacity scaling.}
        \label{fig:simulation_capacity}
    \end{subfigure}
    \hfill
    \begin{subfigure}[t]{0.33\linewidth}
        \centering
        \includegraphics[width=\linewidth]{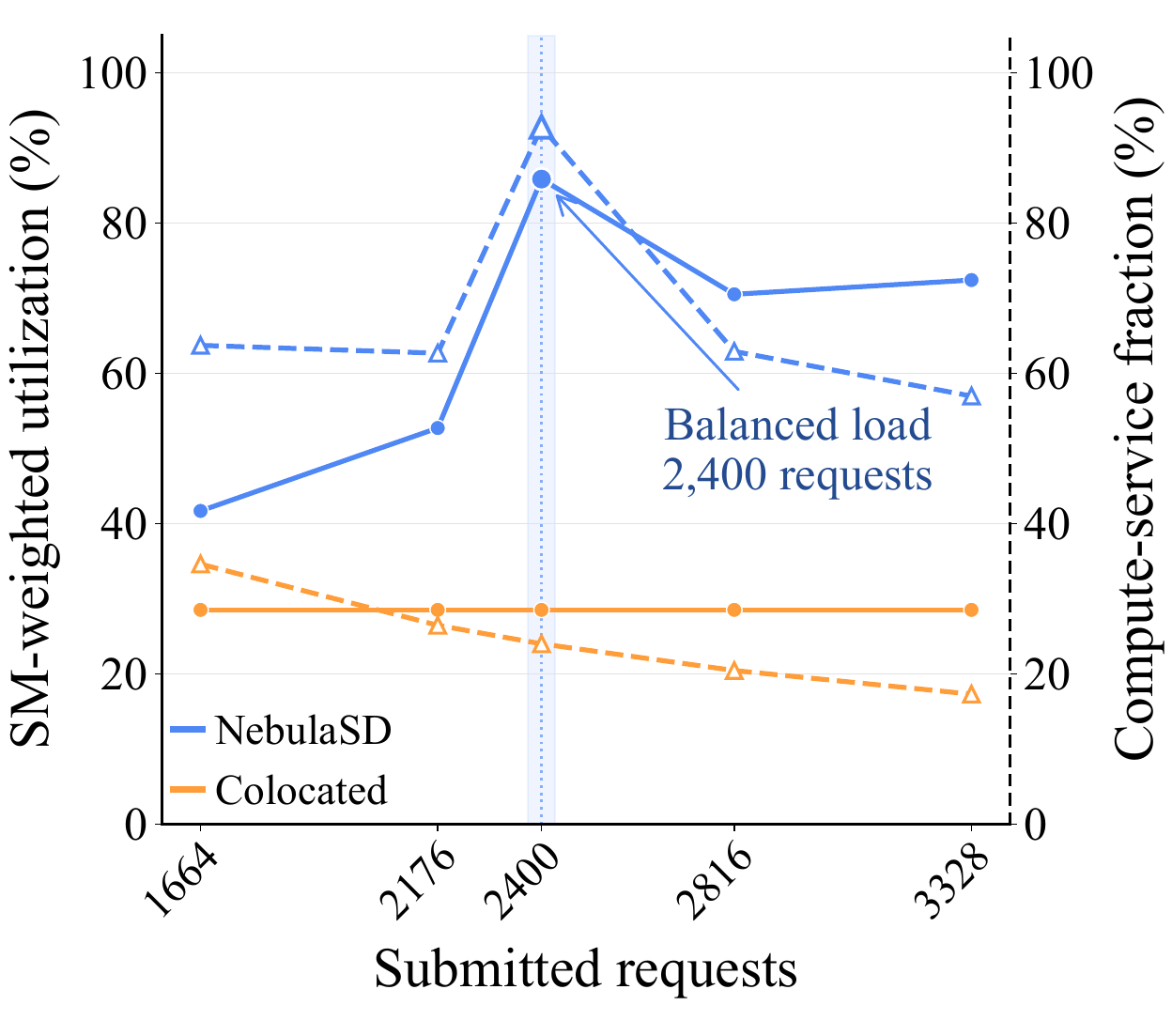}
        \caption{24 GPUs: $4\mathrm{D}+20\mathrm{T}$.}
        \label{fig:simulation_24gpu}
    \end{subfigure}
    \hfill
    \begin{subfigure}[t]{0.32\linewidth}
        \centering
        \includegraphics[width=\linewidth]{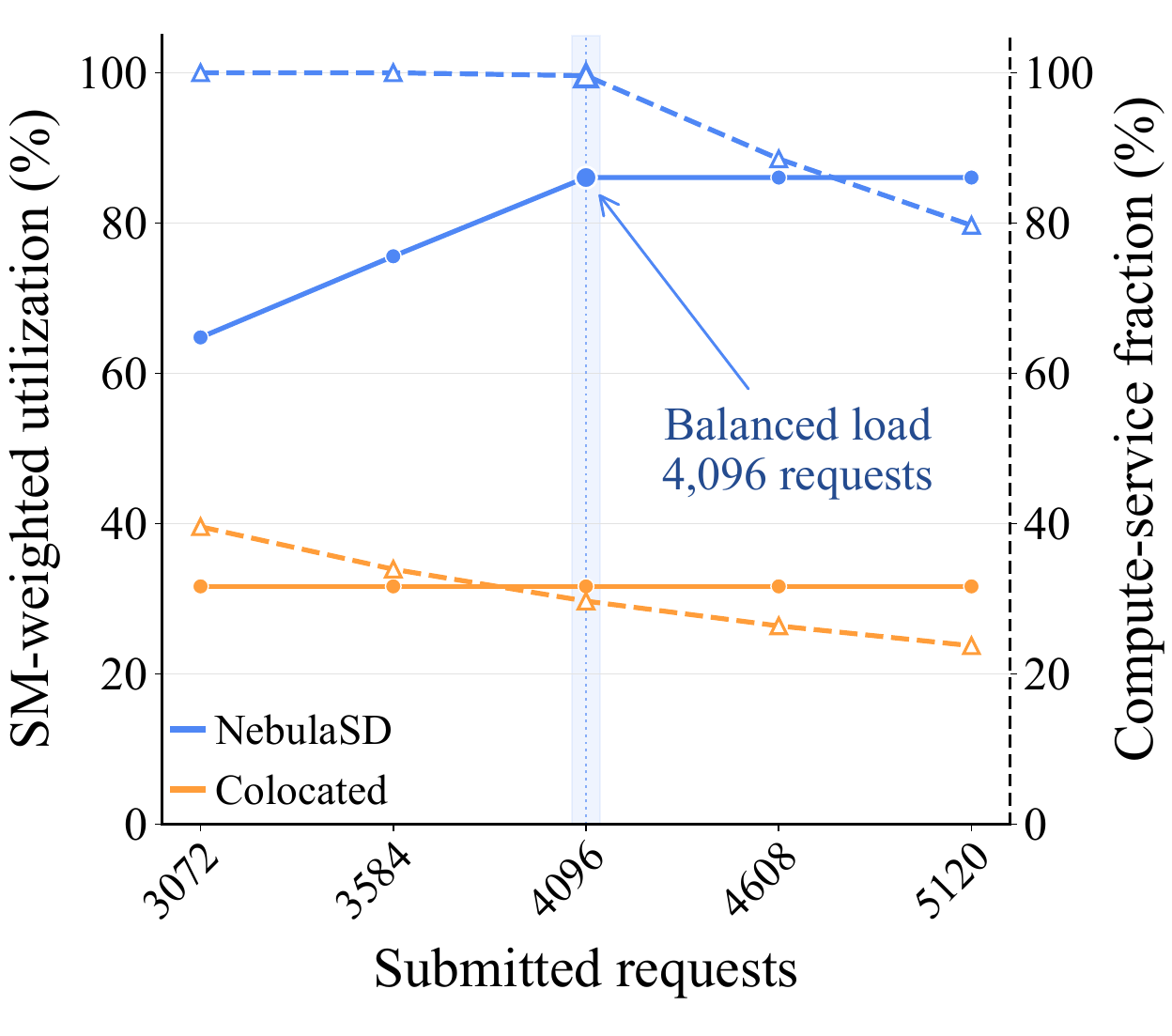}
        \caption{38 GPUs: $6\mathrm{D}+32\mathrm{T}$.}
        \label{fig:simulation_48gpu}
    \end{subfigure}
    \vspace{-5pt}
    \caption{Large-scale simulation of NebulaSD. (a) theoretical capacity and ideal concurrency, (b) and (c) utilization and request service fraction on 24 and 38 GPUs, comparing NebulaSD with Co-located execution.}
    \label{fig:simulation_scaling}
    \vspace{-10pt}
\end{figure}

The four-GPU experiments in Section~\ref{sec:eval_four_gpu} show that dynamic pooling substantially improves hardware utilization and system request-round processing rate, but under a constrained resource budget this comes with a lower per-request compute-service fraction. We therefore use large-scale simulation to examine the compute-side scaling of dynamic pooling under idealized state movement. Simulation details and extended scaling and load-sensitivity analyses are provided in Appendix~\ref{app:simulation}. Fig.~\ref{fig:simulation_capacity} shows that, when the draft and target pools are provisioned according to their relative service capacities, the aggregate processing capacity increases approximately proportionally with system size under this idealized execution model. This suggests that the pooled architecture does not introduce an inherent stage-level scalability bottleneck on the compute side. The capacity curve serves as a reference for resource provisioning instead of scaling the two stages independently, NebulaSD can increase their pool sizes while preserving a balanced draft-target service ratio.

Fig.~\ref{fig:simulation_24gpu} and~\ref{fig:simulation_48gpu} further show that request concurrency determines where the system operates relative to this capacity. At lower concurrency, the worker pools are underutilized and additional requests can increase hardware utilization without immediately sacrificing request service quality. Around the request-matched region, NebulaSD reaches about $86\%$ time-weighted SM activity while maintaining a mean request compute-service fraction of $92.8\%$-$99.6\%$ across the two configurations. Increasing concurrency beyond this region no longer improves hardware utilization and progressively reduces the per-request compute-service fraction as the pooled system becomes oversubscribed. In contrast, Co-located execution is constrained by fixed local draft-target pairs: when active requests exceed the batch capacity these pairs can serve concurrently, only a subset can participate in each execution round while others wait behind their assigned pair. NebulaSD removes this local concurrency constraint by reconstructing batches across the global draft and target pools, sharing available capacity more flexibly. These results complement the small-scale real-system measurements and show that, with appropriately matched draft and target capacity and workload concurrency, NebulaSD can operate near a balanced point that maintains both high hardware utilization and high per-request service quality as the system scales.






\section{Conclusion}
\label{sec:conclusion}

NebulaSD shows that speculative decoding can benefit from treating draft and target execution as independently provisioned and scheduled serving stages instead of fixed execution pairs. By combining shared request pools, worker-triggered batch reconstruction, and migration-aware state preparation, NebulaSD improves effective hardware utilization while keeping most KV-state movement off the execution critical path. Our four-GPU prototype demonstrates that dynamic pooling can translate physical draft-target disaggregation into substantial request-round processing rate gains over both co-located and fixed-pair distributed execution. Profile-driven simulation further shows the compute-side scaling potential of this design under idealized state movement, highlighting the importance of matching draft and target service capacities as worker pools grow. Overall, our results suggest that efficient speculative serving should jointly consider stage-specific resource provisioning, dynamic batch formation, and state movement rather than preserving fixed cross-stage affinities.

\bibliographystyle{iclr2027_conference}
\bibliography{iclr2027_conference}

\newpage

\appendix

\numberwithin{equation}{section}

\numberwithin{equation}{section}

\section{Overview of Supplementary Material}
\label{app:overview}

This appendix provides additional positioning, experimental analysis, implementation details, and scaling results that complement the system design and evaluation presented in the main text. We first position NebulaSD within prior distributed speculative-decoding systems and clarify how its pool-oriented serving abstraction differs from deployment-driven disaggregation. We then characterize the batch-dependent hardware behavior of the draft and target stages, define the time-weighted SM activity metric used in the evaluation, and analyze the interaction between KV migration and foreground model execution. Next, we describe the runtime mechanisms that support frequent state migration and the practical realization of the worker-triggered scheduling abstraction in Section~\ref{sec:system_model}. We further provide additional real-system experiments that examine migration overhead and robustness across KV-state size, proposal depth, scheduler parameters, request concurrency, and draft-model scale. Finally, we describe the large-scale simulator and extend the scaling analysis beyond the configurations reported in the main text.

\begin{enumerate}
    \item \textbf{Positioning within Distributed Speculative Decoding.} We compare NebulaSD with representative distributed speculative-decoding systems and clarify the distinction between deployment-driven disaggregation and the pool-oriented serving problem studied in this work.
    \item \textbf{SM Activity Profiling and Time-Weighted Utilization.} We characterize how SM activity and model execution latency vary with batch size for the draft and target stages, and define the time-weighted SM activity metric used to quantify effective hardware utilization in the main evaluation.
    \item \textbf{Compute-KV Migration Interference.} We study whether overlapping KV-cache transfers with model execution introduces measurable interference to foreground computation. The analysis considers both host-to-device (H2D) and device-to-host (D2H) KV transfers and distinguishes transfer duration from the observed impact on model forward latency.
    \item \textbf{Detailed System Implementation.} We describe the mechanisms used to implement NebulaSD efficiently, including persistent HostKV allocation, double-buffered GPU KV banks, shared-memory communication, event-driven notification, process-isolated worker execution, and bounded H2D submission.
    \item \textbf{Practical Scheduler.} We complement the scheduling model in the main text with runtime mechanisms required for efficient execution, including event-driven scheduling, early planning on standby banks, planning-progress separation, physical feasibility constraints, and the transition from speculative preparation to actual batch execution.
    \item \textbf{Additional Real-System Evaluation.} We evaluate how NebulaSD behaves as KV-state size grows and examine robustness to proposal depth, scheduler parameters, request concurrency, and draft-model scale on the physical four-GPU testbed.
    \item \textbf{Simulation Methodology and Extended Scaling Analysis.} We describe the profiling-driven discrete-event simulator used for large-scale evaluation, clarify its differences from the physical runtime, and provide extended results on scaling behavior, request concurrency, and the utilization-service trade-off across larger worker pools.
\end{enumerate}

\section{Positioning within Distributed Speculative Decoding}
\label{app:distributed_sd_positioning}

Distributed speculative decoding has recently been explored under several deployment settings, but the separation of draft generation and target verification serves different purposes across these systems. Table~\ref{tab:distributed_sd_positioning} summarizes their problem formulations and primary optimization targets. Rather than differing only in implementation details, these systems expose different system bottlenecks after speculative execution is distributed.

\begin{table*}[t]
\centering
\caption{Representative distributed speculative-decoding systems. The separation of draft and target execution serves different deployment goals and exposes different system bottlenecks.}
\label{tab:distributed_sd_positioning}
\small
\setlength{\tabcolsep}{6pt}
\renewcommand{\arraystretch}{1.18}
\begin{tabularx}{\textwidth}{@{}>{\raggedright\arraybackslash}p{0.18\textwidth}>{\raggedright\arraybackslash}p{0.16\textwidth}>{\raggedright\arraybackslash}X>{\raggedright\arraybackslash}X@{}}
\toprule
\textbf{System} & \textbf{Setting} & \textbf{Motivation} & \textbf{System focus} \\
\midrule
Disaggregated SD~\citep{shi2025disaggregated} & Heterogeneous GPUs & Reuse weaker GPUs & Placement and carbon efficiency \\
\addlinespace[2pt]
SLED~\citep{li2025sled} & Edge + server & Exploit edge compute & Shared verification batching \\
\addlinespace[2pt]
WISP~\citep{li2026wisp} & Edge + server & Improve edge-serving efficiency & Dynamic drafting and SLO-aware batching \\
\addlinespace[2pt]
DSD~\citep{yu2025dsd} & Edge-cloud & Adapt to network and batching effects & Adaptive speculation-window control \\
\addlinespace[2pt]
SPECTRE~\citep{xie2026spectre} & Multi-model cloud & Reuse idle serving capacity & Remote drafting and execution overlap \\
\bottomrule
\end{tabularx}
\end{table*}

\paragraph{Deployment-driven disaggregation.} A common motivation in prior distributed speculative-decoding systems is to exploit resources that are difficult to use efficiently under conventional colocated serving. Disaggregated speculative decoding targets heterogeneous accelerator fleets, where a smaller draft model can make productive use of older or lower-end GPUs while target execution remains on more capable devices~\citep{shi2025disaggregated}. In this setting, physical draft-target separation is primarily a mechanism for heterogeneous resource reuse and carbon-efficient deployment. SLED considers a different source of distributed compute: edge devices that would otherwise contribute little to centralized LLM inference. Each device performs lightweight drafting locally, while a shared server executes the target model and batches verification requests arriving from multiple devices~\citep{li2025sled}. WISP builds on this edge-side drafting setting and identifies wasted drafting time and verification interference as its main bottlenecks. Its speculation controller and SLO-aware verification scheduler therefore optimize how much work is generated at the edge and how heterogeneous verification requests are served at the centralized target~\citep{li2026wisp}.

\paragraph{Communication- and resource-driven distributed execution.} Other work focuses on the execution dynamics introduced by distributing speculative decoding itself. DSD studies coordinated draft-target execution across heterogeneous edge-cloud environments, where network latency, batching behavior, and scheduling jointly determine whether speculation remains beneficial. Its primary control variable is the speculation window, which is dynamically adapted to the current execution conditions~\citep{yu2025dsd}. SPECTRE considers multi-model cloud serving rather than edge-cloud deployment. Its motivation comes from long-tailed model demand: lightly loaded model services can provide remote drafting capacity for heavily loaded target services. The resulting scheduling problem is therefore centered on harvesting remote draft capacity while maintaining useful overlap between drafting and target verification under multi-tenant traffic~\citep{xie2026spectre}. Together, these studies show that distributed speculative execution can expose useful resources across heterogeneous GPUs, edge devices, edge-cloud deployments, and multi-model services, while also introducing deployment-specific communication and scheduling challenges.

\paragraph{Pool-oriented speculative serving.} The system problem considered in this work begins from a different point in the design space. We assume that draft and target resources have already been physically separated and ask how those resources should be organized when many concurrent requests repeatedly alternate between the two stages. The physical separation of draft and target execution is therefore a starting point rather than the optimization objective itself. Because drafting and verification can exhibit different batch-size scaling and service capacities, statically preserving request or batch affinity after disaggregation can fragment capacity across workers: one worker or stage may become locally constrained while usable capacity remains elsewhere.

This observation leads to a different system abstraction. Rather than treating distributed speculative decoding primarily as communication between a drafter and a verifier, we model the two stages as independently provisioned worker pools, $\mathcal{D}=\{d_1,\ldots,d_M\}$ and $\mathcal{T}=\{t_1,\ldots,t_N\}$. Requests repeatedly enter the shared request pool of their next stage and need not preserve worker or batch affinity across speculative rounds. The resulting scheduling question is therefore which requests should form the next batch of an available worker, subject to predecessor readiness, worker availability, state readiness, and the service characteristics of that stage. This motivates the worker-triggered batch reconstruction formulation in Section~\ref{sec:system_model}.

Removing worker affinity also changes the role of state movement. In deployment-oriented distributed speculative decoding, communication is primarily induced by separating draft and target execution across devices or services. Under pool-oriented scheduling, state movement additionally arises from dynamically reassigning a request among workers of the \emph{same} stage across successive speculative rounds. NebulaSD therefore treats KV-state migration and advance preparation as enabling mechanisms for resource pooling: state is prepared according to upcoming scheduling decisions and overlapped with foreground execution so that dynamic reassignment does not directly serialize worker computation. Consequently, the central system quantities in our study are stage service capacity, batch formation efficiency, worker utilization, request service, and migration-induced execution gaps, rather than deployment-specific objectives such as edge utilization, carbon efficiency, speculation-window adaptation, or remote-drafter reuse.

\section{SM Activity Profiling and Time-Weighted Utilization}
\label{app:sm_activity}

\paragraph{Time-weighted SM activity.}
Compute-time fraction alone does not fully characterize GPU utilization. A worker may spend a large fraction of time executing model forwards while each forward activates only a small fraction of the available SM resources. Conversely, high SM activity during individual forwards does not imply high system utilization if long idle gaps remain between consecutive computations. We therefore use \emph{time-weighted SM activity} in Section~\ref{sec:eval_four_gpu} to jointly capture how frequently the GPUs perform model computation and how intensively their SM resources are exercised during those computations.

Let $T_{g,k}^{\mathrm{comp}}$ denote the duration of compute interval $k$ on GPU $g$, and let $A_{g,k}\in[0,1]$ denote the calibrated SM Active corresponding to the stage and batch configuration executed during that interval. Over a measurement window of length $T_{\mathrm{window}}$ on $G$ GPUs, we define the system-level time-weighted SM activity as
\begin{equation}
U_{\mathrm{eff}}
=
\frac{
\sum_{g}\sum_{k}
T_{g,k}^{\mathrm{comp}} A_{g,k}
}{
G T_{\mathrm{window}}
}.
\label{eq:time_weighted_sm}
\end{equation}
Idle GPU-time therefore contributes zero, while compute intervals with higher SM activity contribute proportionally more. The contribution of a particular stage is obtained by restricting the summation to compute intervals of that stage, and the draft and target contributions sum to $U_{\mathrm{eff}}$. This metric captures both temporal utilization and within-batch hardware activity. SM Active is obtained from offline profiling rather than sampled during the end-to-end experiments to avoid profiler interference.

\begin{figure}[t]
\centering
\begin{subfigure}[t]{0.49\linewidth}
    \centering
    \includegraphics[width=\linewidth]{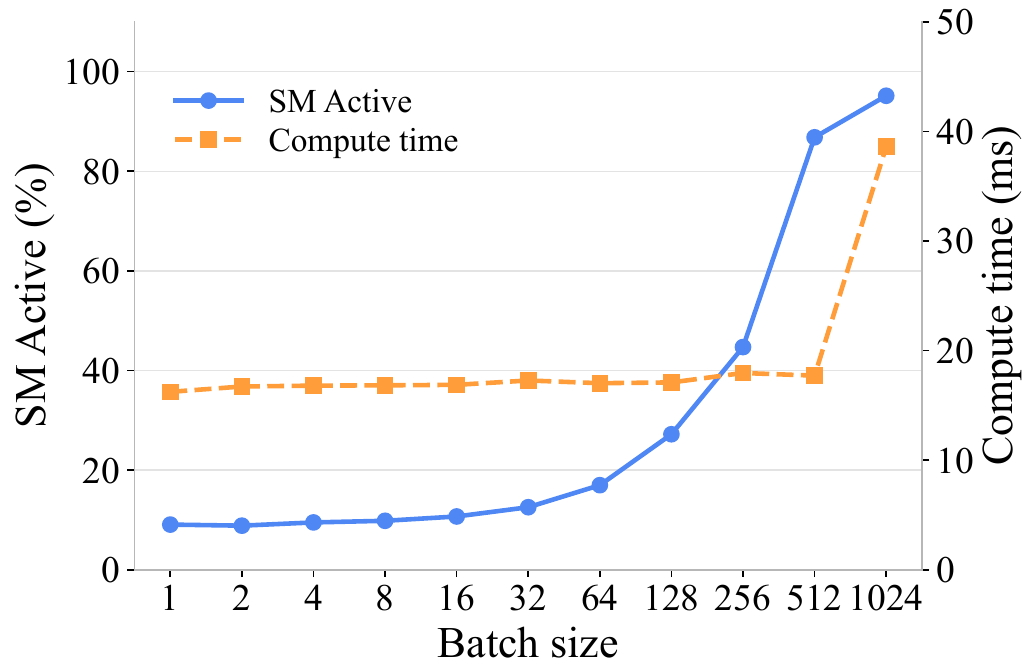}
    \caption{Draft model.}
    \label{fig:sm_draft}
\end{subfigure}
\hfill
\begin{subfigure}[t]{0.49\linewidth}
    \centering
    \includegraphics[width=\linewidth]{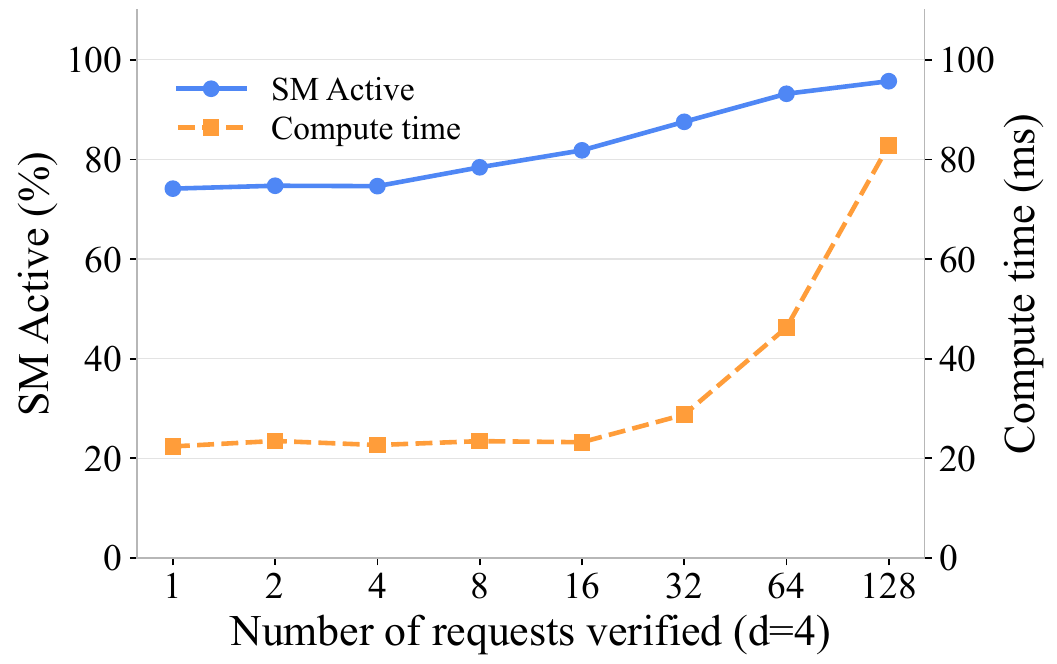}
    \caption{Target model.}
    \label{fig:sm_target}
\end{subfigure}
\caption{Batch-size profiling of SM activity and computation latency under a fixed context length of 128. (a) A single draft-model forward. (b) target verification with speculative depth $d=4$. SM Active is measured during model computation, while computation latency is obtained from runs without SM sampling.}
\label{fig:batch_sm_profile}
\end{figure}

\paragraph{Batch-size profiling.}
Fig.~\ref{fig:batch_sm_profile} shows that the draft and target stages exhibit substantially different relationships between batch size, hardware activity, and compute latency. For the draft model, increasing the batch size from $1$ to $512$ raises SM Active from approximately $9.1\%$ to $86.8\%$, while the latency of a single forward remains roughly within $16$-$18$~ms. Only at batch size $1024$ does the compute latency increase substantially, reaching $38.6$~ms with an SM Active of $95.1\%$. This indicates substantial batching headroom in the draft stage, larger batches can activate considerably more GPU resources without proportionally increasing forward latency.

The target stage reaches high SM activity much earlier. Its SM Active is already approximately $74.1\%$ at verifying $1$ request with $4$ drafting tokens and reaches $87.5\%$ at $32$ requests. Increasing the number of verifying requests further to $64$ and $128$ raises SM Active to approximately $93.2\%$ and $95.7\%$, respectively, but also increases compute latency from approximately $28.8$~ms at batch size $32$ to $46.3$ and $82.8$~ms. The target stage therefore has considerably less batching headroom than the draft stage. These different scaling characteristics motivate stage-specific batch reconstruction in NebulaSD, preserving the same batch organization across the two stages can either leave substantial draft capacity unused or push the target stage into a less favorable latency regime.

\paragraph{Relation to the end-to-end evaluation.}
The time-weighted metric in Eq.~\ref{eq:time_weighted_sm} combines the two utilization dimensions exposed by this profiling. A configuration with high compute occupancy may still have low effective hardware activity if it repeatedly executes poorly utilized batches, while a configuration with high SM Active may remain inefficient if long idle gaps separate consecutive computations. NebulaSD improves both aspects by reconstructing stage-specific batches and reducing execution bubbles between successive computations. This explains why the time-weighted SM activity reported in Section~\ref{sec:eval_four_gpu} provides a more complete view of effective hardware utilization than compute occupancy or average batch size alone.

\section{Compute-KV Migration Interference}
\label{app:kv_interference}

\paragraph{Experimental setup.}
NebulaSD overlaps KV migration with foreground model execution to hide state-transfer latency from the critical path. However, asynchronous H2D and D2H transfers may still interfere with model execution through shared GPU, interconnect, and runtime resources. We therefore conduct a single-GPU microbenchmark to quantify this interference under controlled transfer sizes. The draft experiment uses Qwen3-0.6B with batch size 128, while the target experiment uses Qwen3-8B with batch size 32 and speculative depth $d=4$. For each model, we overlap foreground execution with H2D or D2H KV transfers ranging from 2~MiB to 1024~MiB and compare two implementations, one in which transfer submission shares the foreground execution process and one in which migration is handled by a separate process. H2D follows NebulaSD's bounded submission policy with 32~MiB transfer chunks, while D2H uses the current whole-plan submission path. We report the relative increase in foreground execution latency over the corresponding no-migration baseline,

\begin{equation}
\Delta T
=
\frac{T_{\mathrm{overlap}}-T_{\mathrm{base}}}
{T_{\mathrm{base}}}.
\label{eq:kv_overlap_slowdown}
\end{equation}

\begin{figure}[t]
\centering
\begin{subfigure}[t]{0.49\linewidth}
    \centering
    \includegraphics[width=\linewidth]{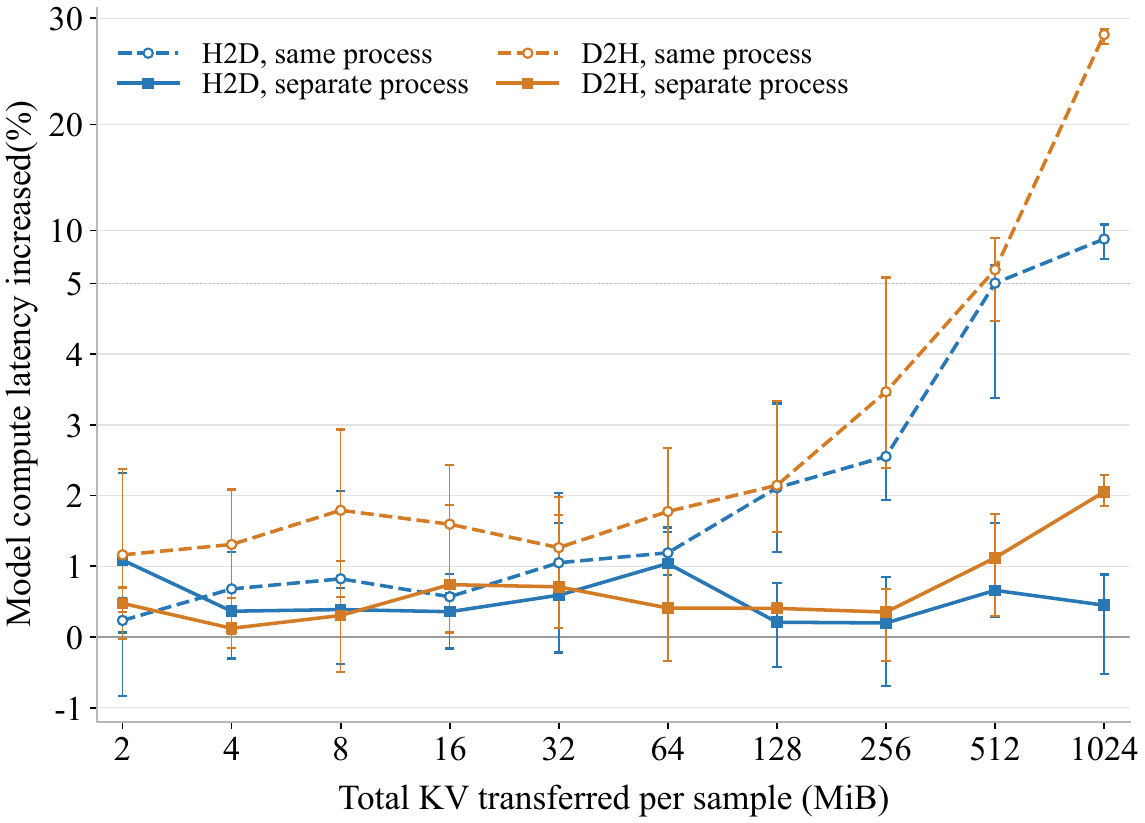}
    \caption{Draft model.}
    \label{fig:kv_overlap_draft}
\end{subfigure}
\hfill
\begin{subfigure}[t]{0.49\linewidth}
    \centering
    \includegraphics[width=\linewidth]{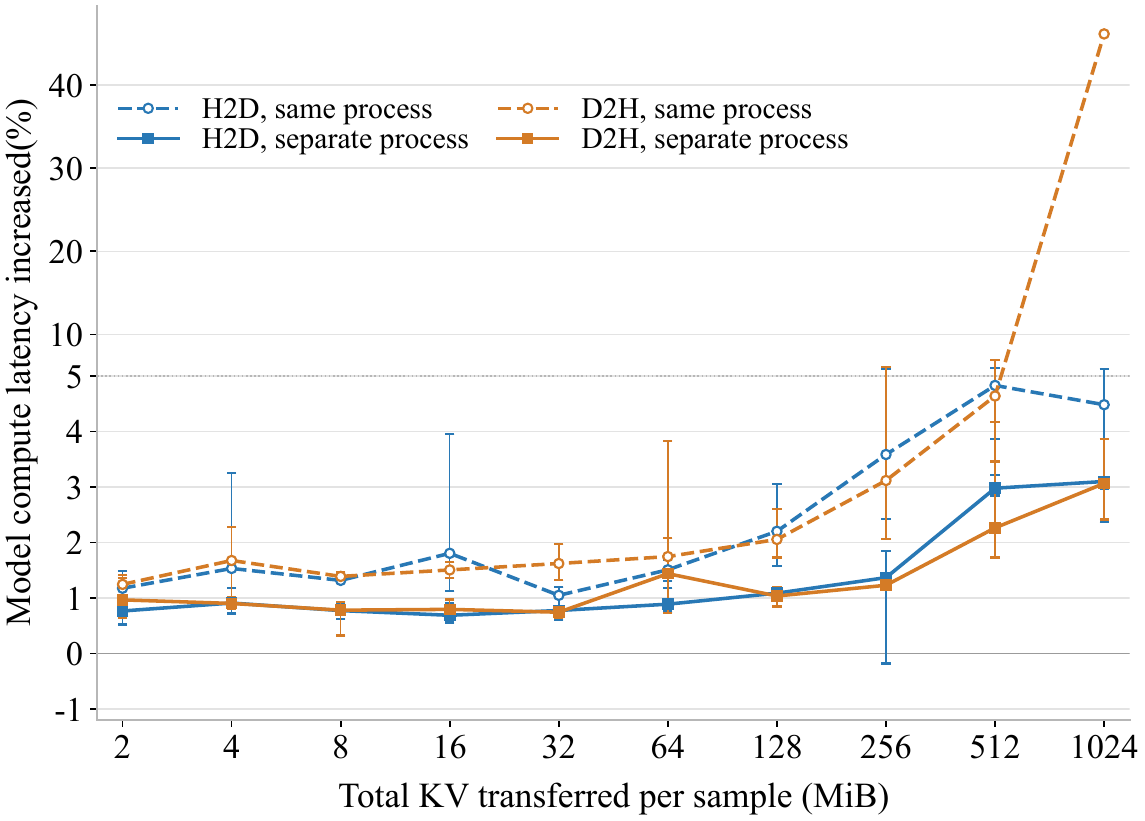}
    \caption{Target model.}
    \label{fig:kv_overlap_target}
\end{subfigure}
\caption{Foreground execution slowdown when KV migration overlaps with model execution. We compare H2D and D2H transfers submitted from the foreground process and from a separate migration process. The y-axis reports latency increase relative to the no-migration baseline; the region above $5\%$ is visually compressed to preserve detail in the low-interference regime.}
\label{fig:kv_overlap}
\end{figure}

\paragraph{Overlap interference.}
Fig.~\ref{fig:kv_overlap} shows that KV migration can generally overlap with foreground computation at low cost for small and moderate transfer sizes, while very large transfers expose stronger interference. The effect is particularly pronounced for same-process D2H submission. At 1024~MiB, the draft execution slowdown reaches $28.47\%$ and the target slowdown reaches $46.13\%$. Moving migration to a separate process reduces these values to only $2.05\%$ and $3.07\%$, respectively. H2D exhibits the same qualitative benefit, although the interference is less severe, at 1024~MiB, process isolation reduces the slowdown from $9.19\%$ to $0.45\%$ for draft execution and from $4.48\%$ to $3.10\%$ for target execution. These results indicate that asynchronous transfer itself is not necessarily expensive for foreground execution. Instead, how large migration operations are submitted and coordinated can determine whether their tail becomes exposed on the foreground critical path.

\paragraph{Source of the large-D2H slowdown.}
The sharp increase observed for large same-process D2H transfers cannot be explained by transfer duration alone. When the D2H size increases from 512~MiB to 1024~MiB, the measured copy-active time grows approximately linearly from $20.4$~ms to $40.8$~ms. In contrast, foreground execution latency increases nonlinearly, the draft slowdown rises from approximately $6.3\%$ to $28.5\%$, while the target slowdown rises from approximately $4.6\%$ to $46.1\%$. Nsight traces further show that the model kernels themselves change only slightly. For draft execution, kernel-active time increases from approximately $26.046$~ms to $26.318$~ms, while the corresponding target kernel-active time changes from approximately $26.824$~ms to $27.225$~ms. The large end-to-end slowdown therefore does not correspond to a proportional slowdown of foreground model kernels.

Instead, the traces show that a latency-sensitive result readback issued by the foreground model can be delayed behind the large background D2H operation. Under same-process 1024~MiB D2H migration, the observed readback wait reaches approximately $15.45$~ms for draft execution and $7.58$~ms for target execution. With migration isolated in a separate process, these waits decrease to approximately $0.66$~ms and $0.26$~ms, respectively. In the same-process traces, the small foreground readback tends to begin only after the tail of the large KV transfer has completed, whereas process isolation allows it to make progress substantially earlier. The evidence therefore suggests that the dominant large-transfer penalty arises from transfer scheduling and head-of-line interference that delays latency-sensitive foreground operations, rather than from a sudden bandwidth collapse or a comparable slowdown of the model kernels themselves.

\paragraph{Implications for NebulaSD.}
These results show that KV migration can be aggressively overlapped with model execution without substantially extending the foreground critical path, provided that large background transfers are prevented from blocking latency-sensitive operations. In particular, the large-D2H experiments show that the dominant exposed overhead arises from transfer scheduling and delayed foreground readback rather than a proportional slowdown of model kernels. This observation motivates NebulaSD's process-isolated migration path and bounded transfer submission, whose implementation is described in Appendix~\ref{app:worker_isolation}. The experiment is a controlled single-GPU microbenchmark and does not uniquely attribute the observed interference to a specific CUDA runtime or copy-engine mechanism. Nevertheless, it demonstrates that careful transfer coordination is necessary for hiding KV migration behind useful computation.

\section{Detailed System Implementation}
\label{app:detailed_implementation}

NebulaSD separates the scheduling abstraction from the mechanisms used to manage KV residency and execute individual batches. Both draft and target workers expose the same runtime interface to the scheduler, while maintaining independent model-specific KV storage. We describe the implementation details that are most relevant to state migration below.

\subsection{Persistent HostKV Allocation}
\label{app:hostkv}

NebulaSD maintains long-lived host-resident KV arenas for both the draft and target models. Each arena is backed by a shared-memory region and contains contiguous K and V planes. Workers address KV state using numeric offsets within these arenas rather than transferring process-local pointers or dynamically allocated buffers.

For a request $r$ with prompt length $P_r$, maximum generation length $G_r$, and KV block size $B$, the Engine reserves an extent of

\begin{equation}
C_r =
\left\lceil
\frac{P_r + G_r}{B}
\right\rceil
\label{eq:app_hostkv_capacity}
\end{equation}

blocks when the request is admitted. The extent remains stable throughout the request lifetime and is recycled only after the request and all outstanding copies have retired. Consequently, migration does not require repeated host-buffer allocation or address remapping as the request moves between workers.

The shared mapping is registered with CUDA by the direct memory access (DMA) process and the registration is reused across subsequent transfers. Registration is therefore performed at the arena level rather than for individual requests or migration operations. The runtime separately tracks the valid KV length and version of each request, allowing a worker to identify the corresponding HostKV extent using compact metadata.

For writeback, NebulaSD transfers only the KV suffix that has changed since the last committed host version. The persistent prefix remains in HostKV and can be reused by subsequent workers. This reduces D2H traffic while preserving a stable host-side state from which either worker pool can reconstruct the required device-resident state.

\subsection{Double-Buffered GPU KV Banks}
\label{app:double_bank}

Each worker maintains two pre-allocated GPU KV banks. The two banks decouple foreground computation from preparation of the next scheduled batch: while one bank is used by the current batch, the other bank can reconstruct its layout and import KV state for a future batch.

Internally, a bank progresses through the following physical states:

\begin{equation}
\textsc{Free}
\rightarrow
\textsc{Filling}
\rightarrow
\textsc{Ready}
\rightarrow
\textsc{Computing}
\rightarrow
\textsc{Exporting}
\rightarrow
\textsc{Free}.
\label{eq:app_bank_lifecycle}
\end{equation}

A bank enters \textsc{Filling} after the runtime reserves its rows and physical KV blocks. It becomes \textsc{Ready} only after the required metadata and H2D state preparation have completed. The runtime may select a \textsc{Ready} bank for computation once the worker becomes available. After computation, any updated KV suffix is asynchronously exported to HostKV before the physical bank is released.

Importantly, bank retirement is determined by physical completion rather than by model-output availability. The result of a batch may therefore become visible to the scheduler while its old bank is still exporting KV state. The other bank can meanwhile become ready and begin the next computation. This permits, for example,

\begin{equation}
\textsc{Compute}_{k+1}
\parallel
\textsc{D2H}_{k},
\qquad
\textsc{Compute}_{k}
\parallel
\textsc{H2D}_{k+1},
\label{eq:app_bank_overlap}
\end{equation}

without allowing a background transfer to overwrite the storage used by an active computation.

The two banks use separate KV block ranges but share a bounded pool of logical batch rows. The scheduler therefore accounts for rows still retained by the other bank when determining whether a new batch can be prepared. This constraint is discussed further in Section~\ref{app:practical_scheduler}.

\subsection{Shared-Memory Control Plane}
\label{app:shared_control_plane}

NebulaSD separates persistent state publication from explicit command delivery. High frequency runtime facts, such as bank state, compute state, KV readiness, and request progress, are published into fixed layout shared-memory tables. The tables are partitioned by writer role, and each partition contains a dedicated row for every active request. Different runtime components therefore update disjoint memory regions even when they operate on the same request. Within a writer partition, a request row has only one authorized writer at a time, request epochs, worker generations, and operation fences prevent a stale or previously assigned worker from publishing into a row after ownership changes. This single-writer organization avoids mutexes on the publication path. Readers obtain consistent snapshots using a lightweight sequence-based publication protocol, allowing the Engine to observe frequently changing worker state without serializing updates through conventional process-level remote procedure call (RPC) messages.

The Engine maintains the coarse-grained canonical state required for global coordination, including request identity and lifecycle, accepted output, HostKV allocation, and issued-work bookkeeping. Fine-grained execution state, including bank ownership, H2D and D2H progress, and local runtime state, remains worker-owned and is exposed to the Engine through the shared observation tables. This separation avoids continuously mirroring the worker runtime inside the centralized scheduler.

Explicit work submission is carried by bounded single-producer, single-consumer shared-memory rings. The corresponding payload contains numeric identities, offsets, epochs, and compact descriptors rather than KV tensors or Python objects. State-change rings similarly carry only references to modified table rows; the table itself remains the source of truth.

On Linux, NebulaSD uses \texttt{eventfd} as a lightweight doorbell between processes. A doorbell indicates that new work or state changes may be available, but does not carry the corresponding payload. Multiple wakeups may therefore be coalesced without affecting correctness. After receiving a doorbell, the consumer drains the corresponding ring and reads the authoritative shared-memory state. If a notification ring overflows, the reader can recover by scanning publication sequence numbers in the underlying table.

\subsection{Process-Isolated Worker Runtime}
\label{app:worker_isolation}

A logical NebulaSD worker is implemented using three cooperating processes: a \emph{control process}, an \emph{execution process}, and a \emph{DMA process}. The same organization is used for both draft and target workers.

The control process owns the local runtime state machine. It processes scheduler commands, prepares model inputs, manages bank transitions and dependencies, and publishes execution facts to the shared observation plane. The execution process owns the model and its device-resident KV storage and is kept deliberately narrow, once a complete model job is provided by the control process, it performs the corresponding GPU computation and returns the result. In particular, control-plane processing, KV-copy completion handling, and metadata publication do not execute alongside the model in this process.

The DMA process independently owns KV-copy submission and completion. GPU KV storage is shared with the DMA process at worker initialization, after which the control path communicates copy plans using lightweight descriptors. H2D and D2H operations are issued on dedicated CUDA copy streams without requiring the execution process to poll their completion.

This organization is motivated by the latency sensitivity of GPU submission from the host. CPU scheduling delays, Python runtime activity, copy-completion handling, and control-plane work can otherwise introduce gaps between model operations even when sufficient GPU resources are available. Process isolation does not remove these costs, but prevents most of them from sharing the execution process's latency-critical CPU path.

\subsection{Bounded H2D Submission}
\label{app:chunked_h2d}

\begin{figure}[t]
    \centering
    \includegraphics[width=0.8\linewidth]{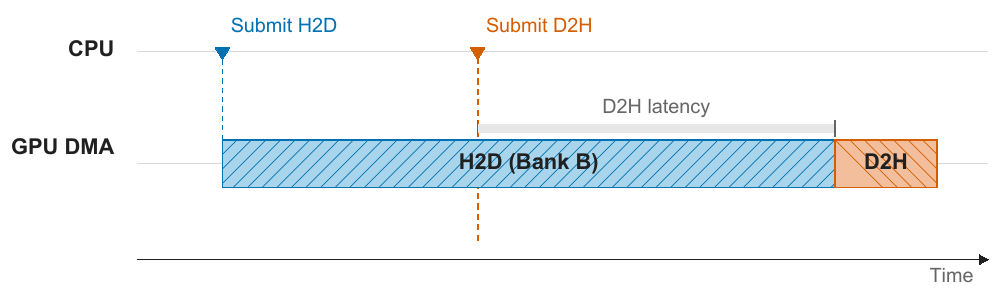}
    \caption{Timeline without chunked H2D transfers.}
    \label{fig:whole-h2d}
\end{figure}

\begin{figure}[t]
    \centering
    \includegraphics[width=0.8\linewidth]{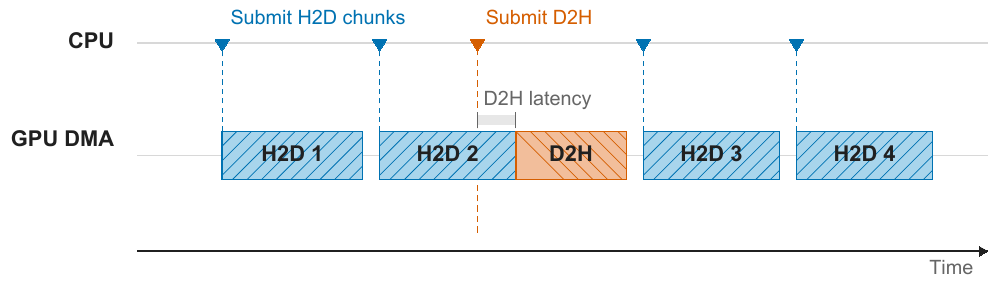}
    \caption{Timeline with chunked H2D transfers.}
    \label{fig:chunked-h2d}
\end{figure}

NebulaSD additionally bounds the amount of H2D work submitted to the GPU at once. Within a single KV bank, H2D preparation and D2H writeback are serialized by the bank lifecycle and therefore do not overlap with each other. However, the two logical banks of a worker can be in different phases simultaneously. For example, while one bank is restoring the KV state of an upcoming batch through H2D transfers, the other bank may still be writing back the updated KV state of the previous batch through D2H transfers. These banks are logical partitions of the same worker GPU rather than independent devices, so their transfers ultimately share the same underlying GPU copy resources.

Although CUDA-capable devices may execute H2D and D2H transfers concurrently, asynchronous submission from different streams does not by itself guarantee that the resulting transfers will overlap efficiently at the device level. Transfers issued for different banks can still contend for the same copy engines and hardware work queues, and a sufficiently long sequence of outstanding H2D operations may delay a D2H transfer issued by the other bank as shown in Fig.~\ref{fig:whole-h2d}. This situation arises naturally in NebulaSD when the standby bank begins restoring a large future batch while the previously active bank is still completing its KV writeback.

To limit this queueing effect, NebulaSD divides large H2D plans into bounded chunks as shown in Fig.~\ref{fig:chunked-h2d}. In our implementation, the default H2D window is 32\,MiB. Only a bounded group of chunks is submitted before the DMA process waits for its completion and submits the next group. D2H writeback from the other bank remains independently submittable during this process. Chunking therefore does not change the per-bank ordering of H2D and D2H operations. Instead, it limits the amount of H2D work from one bank that can accumulate ahead of D2H traffic from the other bank, reducing cross-bank head-of-line blocking on the shared GPU copy path.

This mechanism addresses interference between concurrent KV transfers. Its purpose is distinct from the compute-transfer interference studied in Appendix~\ref{app:kv_interference}, where we measure whether concurrent H2D or D2H activity changes the execution time of foreground model kernels.

\section{Practical Scheduler}
\label{app:practical_scheduler}

The optimization in Section~\ref{sec:system_model} describes worker-triggered batch reconstruction over a set of stage-eligible requests, where the destination worker is determined by the scheduling opportunity. The runtime implementation preserves this abstraction but introduces additional mechanisms to bound scheduling overhead, expose preparation opportunities early, and respect transient physical-resource constraints.

\subsection{Event-Driven Scheduling}
\label{app:event_scheduler}

NebulaSD adopts an event-driven scheduling design rather than periodically rerunning the scheduler at a fixed clock interval. A new scheduling decision is useful only when the runtime state changes in a way that may affect the set of requests available for reconstruction or the preparation capacity exposed by a worker. Repeatedly invoking the scheduler while neither condition has changed would introduce additional control-plane overhead without creating a meaningful new scheduling opportunity.

Accordingly, NebulaSD opens or revisits a worker-local planning opportunity when runtime events change either its candidate requests or its available preparation capacity. Typical examples include worker initialization, request admission, completion or dispatch of an ongoing computation, request cancellation, resource retirement, and bank-state transitions that expose preparation capacity. Multiple events occurring within a short interval are incorporated into a common state snapshot before scheduling, avoiding repeated reconstruction over nearly identical states. To bound scheduling overhead, detailed prediction and reconstruction are restricted to a bounded priority-ordered subset of eligible requests, this implementation bound does not change the worker-local objective in Eq.~\ref{eq:batch_reconstruction}.

\paragraph{Early planning across speculative stages.} NebulaSD may include a request in next-stage reconstruction before its predecessor computation has completed. For example, once a draft batch has been issued, its requests are known to subsequently require target execution, and their predecessor completion times can be predicted from the running draft batch. When a target worker exposes a preparation opportunity on its standby bank, the scheduler may therefore consider such requests for target-side reconstruction before their proposal tokens are available. Their target-side KV states correspond to previously committed tokens and can be prepared as soon as the required HostKV source state is available, while the eventual target execution remains gated by publication of the draft result. The same mechanism applies symmetrically from target to draft.

\paragraph{Early planning on the standby bank.} A preparation opportunity may also be exposed by the worker itself. Once a prepared batch is dispatched for computation, the worker's current compute capacity is occupied, but its other bank may become available for future preparation. If the standby bank satisfies the physical-capacity constraints described in Section~\ref{app:scheduler_constraints}, NebulaSD may immediately reconstruct another future batch for this worker rather than waiting for the running computation to complete. NebulaSD can therefore reconstruct and prepare the worker's next batch on its standby bank while the current batch is still executing.

Together, these mechanisms provide the information and capacity required for look-ahead scheduling: running predecessor batches expose requests that can be considered before their outputs are available, while standby banks expose worker-local preparation opportunities that can host future batches. This allows scheduling and KV-state preparation to overlap with ongoing model execution without changing the execution dependencies between draft and target stages.

\paragraph{Scheduling-execution overlap.} These mechanisms allow both scheduling and state preparation to be moved away from the critical path between consecutive GPU computations. Conceptually, NebulaSD attempts to realize
\begin{equation}
\underbrace{\mathrm{Run}(\mathcal{B}_{k})}_{\text{current computation}}
\quad\parallel\quad
\underbrace{\mathrm{Plan}(\mathcal{B}_{k+1})+\mathrm{Prepare}(\mathcal{B}_{k+1})}_{\text{future batch}},
\end{equation}
rather than strictly serializing
\begin{equation}
\mathrm{Run}(\mathcal{B}_{k})
\rightarrow
\mathrm{Plan}(\mathcal{B}_{k+1})
\rightarrow
\mathrm{Prepare}(\mathcal{B}_{k+1})
\rightarrow
\mathrm{Run}(\mathcal{B}_{k+1}).
\end{equation}

The objective is not simply to invoke the scheduler earlier, but to complete as much future planning and preparation as possible before the current computation retires. When a future batch has already been reconstructed, assigned, and prepared, the completion of the current batch requires only the remaining readiness conditions to be checked before the next computation can start.

\paragraph{Example.} The following example illustrates how this overlap is realized for a target worker with two logical banks. Consider a target worker $t_0$ with two logical banks, $A$ and $B$. Suppose bank $A$ is currently executing target batch $\mathcal{B}_{X}$, while bank $B$ exposes a preparation opportunity for $t_0$. At the same time, draft worker $d_0$ is executing a draft batch $\mathcal{B}_{\mathrm{D}}=\{r_1,r_2,r_3,r_4,r_5,r_6\}$. Although its proposal tokens are not yet available, the predicted completion times of these requests allow them to be considered for future target execution.

For the preparation opportunity on bank $B$ of $t_0$, suppose the target scheduler reconstructs $\mathcal{B}_{Y}=\{r_1,r_2,r_3,r_4\}$ from the shared target request pool. NebulaSD does not require the target stage to preserve the draft batch membership, so $r_5$ and $r_6$ remain available for subsequent worker-triggered reconstruction opportunities. In general, a reconstructed batch may merge requests originating from different predecessor batches or split requests that previously executed together.

Although $\mathcal{B}_{Y}$ cannot yet execute, its batch membership and destination bank are now fixed. The target-side KV states required by $r_1,\ldots,r_4$ correspond to previously committed tokens and therefore do not depend on the proposal currently being generated by the draft model. If these states are available in HostKV, NebulaSD can begin preparing $\mathcal{B}_{Y}$ on bank $B$ while the draft computation is still running and bank $A$ continues executing $\mathcal{B}_{X}$.

Conceptually, one possible execution order is:

\begin{verbatim}
time  -------------------------------------------------------->

draft d0:
      |-------------- draft(B_D) ---------------|
                                                ^
                                                |
                                           InputReady

target t0 / Bank A:
      |--------- target(B_X) ---------|
                                      ^
                                      |
                                 WorkerReady

target t0 / Bank B:
      |--- Prepare(B_Y) ----|...................|-- target(B_Y) --|
                            ^
                            |
                         KVReady
\end{verbatim}

Here, the dotted interval denotes a prepared batch waiting for its remaining dependencies. In the illustrated ordering, the H2D preparation of $\mathcal{B}_{Y}$ completes first, so the required target KV state is already resident on bank $B$. Bank $A$ subsequently finishes $\mathcal{B}_{X}$, restoring target compute capacity. The batch still cannot execute until the draft computation publishes the proposal tokens required by $r_1,\ldots,r_4$. Once the readiness conditions defined in Section~\ref{sec:system_model} are satisfied, $\mathcal{B}_{Y}$ can transition directly from the prepared state to target execution.

Importantly, the completion of KV preparation, the retirement of $\mathcal{B}_{X}$, and the publication of the draft result do not independently trigger reconstruction of $\mathcal{B}_{Y}$. Its membership and destination bank were fixed during the earlier worker-local planning step. These later events only advance the prepared batch toward execution, allowing the progress path to remain lightweight.

\paragraph{Planning and progress paths.} NebulaSD therefore separates two logically distinct control paths. The \emph{planning path} reacts to events that expose or change a worker-local batch-reconstruction opportunity, such as newly available candidate requests or preparation capacity on a standby bank, and invokes the stage scheduler for that worker. The \emph{progress path} handles events that only change the readiness of an already selected batch, such as H2D completion, predecessor-result publication, or restoration of compute capacity. These events perform a lightweight readiness check rather than rerunning batch reconstruction. This separation avoids repeatedly executing the more expensive scheduling logic for a frozen batch while still exposing preparation opportunities early enough to overlap scheduling, KV preparation, and GPU execution.

\subsection{Physical Feasibility Constraints}
\label{app:scheduler_constraints}

Before admitting a reconstructed batch for a triggered worker, the implementation checks whether the batch can be physically prepared on the worker's available bank. In addition to the logical constraints in Section~\ref{sec:system_model}, a candidate batch must satisfy the worker's batch-row capacity, GPU KV-block capacity, and stage-specific token limit. Each request must also fit the worker individually.

Double buffering introduces an additional transient constraint. The two banks share the worker's finite row namespace, and the previously active bank may retain rows while computation or D2H writeback is still in progress. The effective row capacity of the standby bank is therefore reduced by the rows still owned by the other bank. A bank is considered a valid preparation destination only when its current role and physical state permit a new layout to be installed.

The scheduler also treats already issued work as a reservation. A worker with a prepared future batch cannot receive another future preparation for the same destination capacity, and requests already included in an issued preparation are excluded from subsequent reconstruction. These constraints prevent speculative scheduling decisions from overcommitting physical bank resources.

\subsection{From Prepare to Run}
\label{app:prepare_run}

The implementation separates worker capacity for \emph{preparation} from capacity for \emph{computation}. A worker may therefore be unavailable for immediate execution while still being a valid destination for preparation on its standby bank.

Once the scheduler reconstructs a batch for such a preparation opportunity, NebulaSD issues a \textsc{Prepare} operation that fixes the batch membership, destination worker, bank identity, and expected state versions. The worker can then allocate the bank layout and start the required H2D transfer while its current batch is still executing. This prepared batch remains reserved rather than being reconstructed at every subsequent event.

A prepared batch is converted into a \textsc{Run} operation only when the runtime confirms the three conditions represented abstractly in the main text:

\begin{equation}
\textsc{InputReady}
\;\land\;
\textsc{KVReady}
\;\land\;
\textsc{WorkerReady}.
\label{eq:app_ready_join}
\end{equation}

Here, \textsc{InputReady} confirms that the required predecessor-stage result has been published, \textsc{KVReady} confirms that the expected KV version is resident in the assigned bank, and \textsc{WorkerReady} confirms that the worker has regained compute capacity. Runtime identities and epochs are checked again before dispatch so that stale preparations are not executed after cancellation or state changes.

This two-step \textsc{Prepare}$\rightarrow$\textsc{Run} protocol is the runtime realization of the advance-planning model in Section~\ref{sec:system_model}. Predicted completion times determine which future state should be prepared, whereas real runtime events determine when that state is safe to execute.

\section{Additional Real-System Evaluation}
\label{app:real_system_eval}

This section provides additional real-system evaluation of NebulaSD on the four RTX4090 GPUs, focusing on both runtime overhead and robustness across different operating conditions. We first examine how increasing KV-state size affects migration overhead and the extent to which state movement becomes exposed on the execution critical path. We then evaluate sensitivity to the proposal depth and scheduler parameters, vary request concurrency without workload-specific retuning to characterize the operating regime of resource pooling, and replace the default draft model with a larger model to examine robustness to the draft-target capacity balance.

\subsection{Sensitivity to KV-State Growth}
\label{app:kv_growth}

Dynamic request reassignment requires the execution state of a request to follow its scheduling decisions, making migration increasingly important as the KV state grows. We therefore vary the maximum generation length among 128, 256, and 512 tokens while keeping the model pair, 512 submitted requests, proposal depth of 4, scheduler configuration, KV capacity, and worker initialization unchanged. NebulaSD uses two draft and two target workers with maximum batch sizes of 128 and 32, respectively. All three NebulaSD measurements use the same production implementation and differ only in the maximum generation length during formal measurement.

\begin{figure}[t]
\centering
\includegraphics[width=0.8\linewidth]{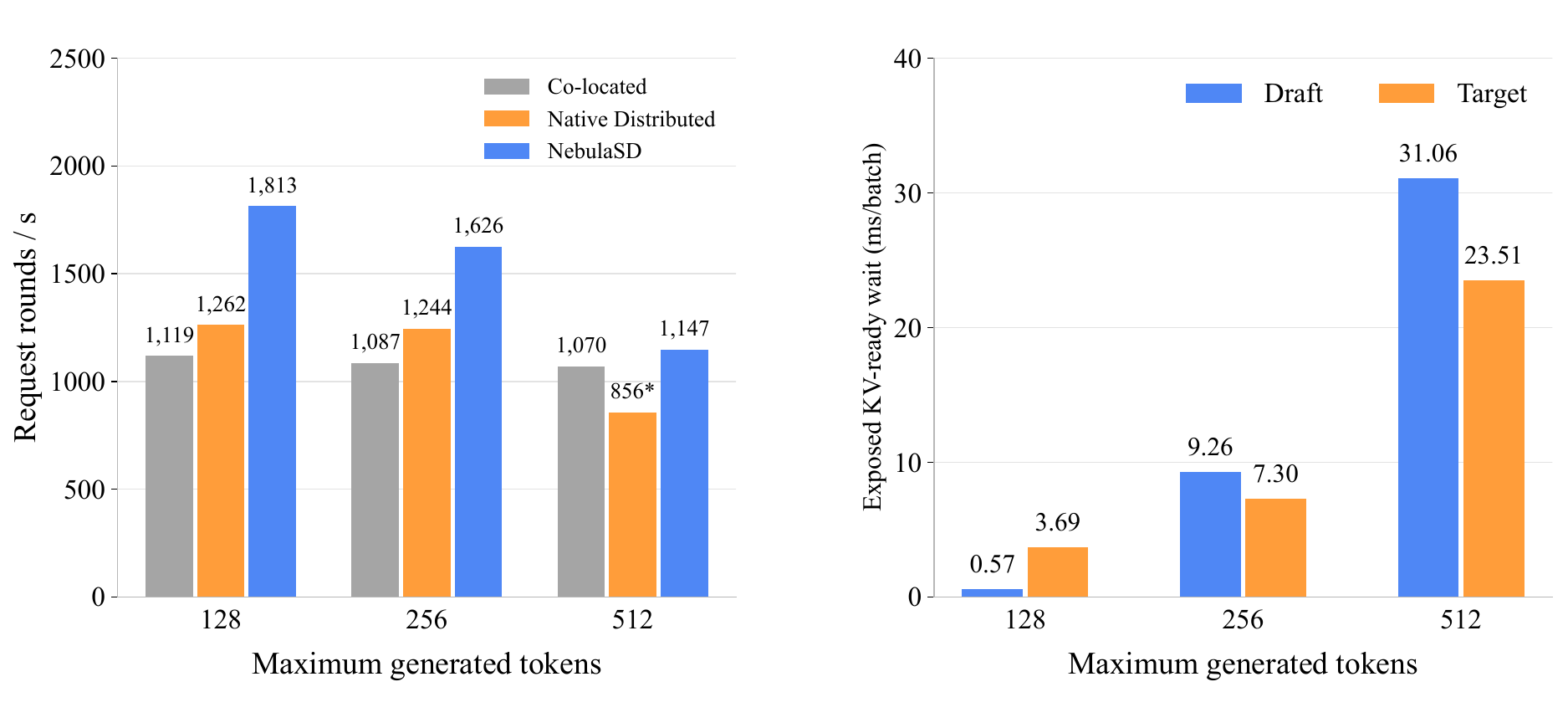}
\caption{Sensitivity to KV-state growth. Left: request-round processing rate as the maximum generation length increases. Right: exposed KV-ready waiting time per batch in NebulaSD. The Native Distributed result at 512 tokens (\textsuperscript{*}) uses 64 rather than 128 resident requests per draft-target pair because the original configuration exceeds target-GPU memory capacity.}
\label{fig:kv_growth}
\end{figure}

Figure~\ref{fig:kv_growth} shows that the benefit of dynamic pooling decreases as the KV state grows. At maximum generation lengths of 128 and 256 tokens, NebulaSD processes 1813.28 and 1625.96 request rounds/s, compared with 1118.62 and 1086.59 rounds/s for Co-located and 1262.24 and 1244.47 rounds/s for Native Distributed. This corresponds to improvements of 62.1\% and 49.6\% over Co-located and 43.7\% and 30.7\% over Native Distributed, respectively. At 512 tokens, NebulaSD reaches 1146.78 rounds/s, remaining 7.1\% above the 1070.41 rounds/s of Co-located execution. The original Native Distributed configuration cannot retain 128 requests per pair at this length because its target-side resident KV state exceeds the available GPU-memory capacity, the capacity-adjusted run shown in Fig.~\ref{fig:kv_growth} therefore uses 64 requests per pair and is not treated as an equal-configuration performance comparison. All methods receive the same 512 submitted requests, while their resident active concurrency follows the corresponding serving architecture.

\paragraph{Exposed migration delay.} Raw transfer latency does not directly determine execution overhead because KV migration can overlap with predecessor execution and worker computation. We therefore measure only the portion of KV preparation that remains exposed after the other execution dependencies are ready. For batch $i$, let $I_i$ denote its input-ready time, $W_i$ the time when the destination worker becomes available, $H_i$ the completion time of its H2D migration, and $M_i$ the completion time of KV metadata preparation. Its KV-ready time is $K_i=\max(H_i,M_i)$, and we define the exposed KV-ready wait as
\begin{equation}
E_i=\left[K_i-\max(I_i,W_i)\right]_+.
\label{eq:kv_exposed_wait}
\end{equation}
This metric counts only the interval during which input and worker readiness would otherwise permit execution but KV preparation remains incomplete; migration hidden behind another unresolved dependency therefore contributes zero exposed wait. To isolate the migration component further, we additionally compute $E_i^{\mathrm{H2D}}=[H_i-\max(I_i,W_i,M_i)]_+$.

\begin{table}[t]
\centering
\caption{KV-migration statistics under increasing KV-state size. Values are draft / target. Effective H2D rate is measured over the observed migration-execution interval and should not be interpreted as the physical PCIe peak bandwidth.}
\label{tab:kv_growth_migration}
\small
\setlength{\tabcolsep}{6pt}
\renewcommand{\arraystretch}{1.12}
\begin{tabular}{@{}cccc@{}}
\toprule
\textbf{Max gen.} & \textbf{H2D/batch (MiB)} & \textbf{Effective H2D (GB/s)} & \textbf{Exposed KV wait (ms)} \\
\midrule
128 & 690.65 / 307.50 & 19.91 / 18.65 & 0.57 / 3.69 \\
256 & 1065.31 / 491.60 & 16.72 / 17.34 & 9.26 / 7.30 \\
512 & 1837.07 / 911.24 & 13.42 / 14.34 & 31.06 / 23.51 \\
\bottomrule
\end{tabular}
\end{table}

Table~\ref{tab:kv_growth_migration} shows that KV migration becomes increasingly exposed as the request state grows. Larger KV states increase the amount of data transferred per batch and keep transfers in flight for longer, while the observed effective H2D rate decreases as concurrent transfers increasingly contend for the shared host-device transfer path. Together, the larger migration volume and lower effective transfer rate cause more KV preparation to remain unfinished after input and worker readiness have been satisfied, increasing the exposed KV-ready wait at longer generation lengths. The H2D-only exposed-wait estimate closely tracks the overall KV-ready wait across all evaluated lengths and both stages, differing by at most 0.49 ms per batch, indicating that H2D migration rather than metadata preparation is the dominant source of exposed KV delay.

This behavior primarily reflects the host-device bandwidth limitation of our four-RTX-4090-GPU testbed. NebulaSD continues to overlap KV migration with model execution using the same scheduling and preparation pipeline, but the KV footprint grows with sequence length while the available migration bandwidth remains fixed. Longer transfers also increase contention among concurrent H2D and D2H operations on the shared PCIe and GPU copy path. As a result, an increasing fraction of otherwise asynchronous KV migration becomes exposed on the execution critical path, which is the primary cause of the processing-rate degradation observed as the maximum generation length increases from 128 to 512 tokens. Because the reported effective H2D rate includes chunk submission, synchronization, and concurrent use of the copy path, we do not attribute its reduction solely to the physical PCIe link; rather, the results identify the available host-device transfer capacity of the current platform as the main bottleneck limiting migration overlap in this regime.

\subsection{Sensitivity to Proposal Depth and Scheduler Parameters}
\label{app:parameter_sensitivity}

We next examine whether the measured benefit depends on the default proposal depth or on a narrowly tuned scheduler configuration. These experiments use 512 submitted requests with a 128-token generation limit. Each condition is repeated three times, and Table~\ref{tab:real_parameter_sensitivity} reports the mean request-round processing rate and one sample standard deviation. For scheduler sensitivity, the proposal depth remains fixed at 4 and only one parameter group is varied at a time.

\begin{table}[t]
\centering
\caption{Real-system sensitivity to the proposal depth and scheduler parameters. Processing rates are completed request rounds/s and are reported as mean $\pm$ one sample standard deviation over three runs.}
\label{tab:real_parameter_sensitivity}
\small

\begin{subtable}[t]{\linewidth}
\centering
\caption{Proposal-depth sensitivity.}
\begin{tabular}{@{}crrrrr@{}}
\toprule
\textbf{Depth} & \textbf{Co-located} & \textbf{Native} & \textbf{NebulaSD} & \textbf{Gain vs. Co-loc.} & \textbf{Gain vs. Native} \\
\midrule
3 & $1359\pm8$ & $1506\pm3$ & $1968\pm42$ & $+44.8\%$ & $+30.7\%$ \\
4 & $1117\pm0$ & $1269\pm4$ & $1884\pm25$ & $+68.7\%$ & $+48.5\%$ \\
5 & $976\pm3$ & $1153\pm3$ & $1682\pm57$ & $+72.3\%$ & $+45.9\%$ \\
\bottomrule
\end{tabular}
\end{subtable}

\medskip

\begin{subtable}[t]{\linewidth}
\centering
\caption{Scheduler-parameter sensitivity.}
\begin{tabular}{@{}crrcrr@{}}
\toprule
\multicolumn{3}{c}{$g_s$ scale} & \multicolumn{3}{c}{$\Delta_s$} \\
\cmidrule(lr){1-3}\cmidrule(lr){4-6}
\textbf{Setting} & \textbf{Rounds/s} & \textbf{$\Delta$} & \textbf{Setting} & \textbf{Rounds/s} & \textbf{$\Delta$} \\
\midrule
$0.5\times$ & $1864\pm12$ & $-1.05\%$ & 0 ms & $1890\pm9$ & $+0.29\%$ \\
$1.0\times$ & $1884\pm25$ & -- & 30 ms & $1884\pm25$ & -- \\
$1.5\times$ & $1887\pm10$ & $+0.17\%$ & 60 ms & $1888\pm23$ & $+0.22\%$ \\
\bottomrule
\end{tabular}
\end{subtable}

\end{table}

\paragraph{Proposal depth.} The main evaluation uses a proposal depth of 4. We additionally evaluate depths of 3 and 5 while keeping the submitted workload and resource configuration unchanged. Changing the proposal depth changes the computation performed by both stages in each speculative round and therefore changes the absolute request-round rate of all three systems. Nevertheless, NebulaSD consistently provides the highest processing rate: its rates at depths 3, 4, and 5 are $1.31\times$, $1.48\times$, and $1.46\times$ those of Native Distributed and $1.45\times$, $1.69\times$, and $1.72\times$ those of Co-located execution, respectively. The resource-pooling benefit is therefore not specific to the default depth of 4.

For depths 3 and 5, the scheduler retains the measured depth-4 profiling table and scales predicted compute cost according to the executed proposal workload: draft cost is scaled with the number of draft forwards and target cost with the anchor-plus-proposal token count. This scaling affects only online cost prediction; all processing rates in Table~\ref{tab:real_parameter_sensitivity} are measured from real model execution. We therefore use these experiments to test robustness of the serving architecture across nearby proposal depths rather than to characterize proposal-depth selection itself.

\paragraph{Scheduler parameters.} The reconstruction policy exposes the stage-specific target service interval $g_s$ and the additional batch-formation slack $\Delta_s$. The default draft/target service intervals are 160/130 ms and the default slack is 30 ms. We jointly scale the two service intervals by $0.5\times$, $1.0\times$, and $1.5\times$, corresponding to 80/65, 160/130, and 240/195 ms, and separately vary $\Delta_s$ among 0, 30, and 60 ms while keeping $g_s$ at its default. The request-round rate remains within 1.05\% of the default across all evaluated configurations. The measured performance therefore does not rely on a narrow optimum of either scheduler parameter within the evaluated range.

\subsection{Sensitivity to Request Concurrency}
\label{app:request_concurrency}

We further evaluate whether the benefit of NebulaSD depends on the 512-request workload used in the main evaluation. We vary the number of simultaneously submitted requests among 256, 384, and 512 while keeping the model pair, the $2\mathrm{D}+2\mathrm{T}$ resource configuration, the proposal depth, the batch-size limits, and the scheduler parameters unchanged. In particular, the draft/target target service intervals remain 160/130 ms and the batch-formation slack remains 30 ms for all three workloads, without workload-specific retuning. The resident baselines retain their original execution organizations, whereas NebulaSD admits the submitted requests into shared stage-level request pools. This experiment therefore evaluates how the original serving architectures respond to increasing offered concurrency rather than enforcing equal resident concurrency across methods.

\begin{figure}[t]
\centering
\includegraphics[width=\linewidth]{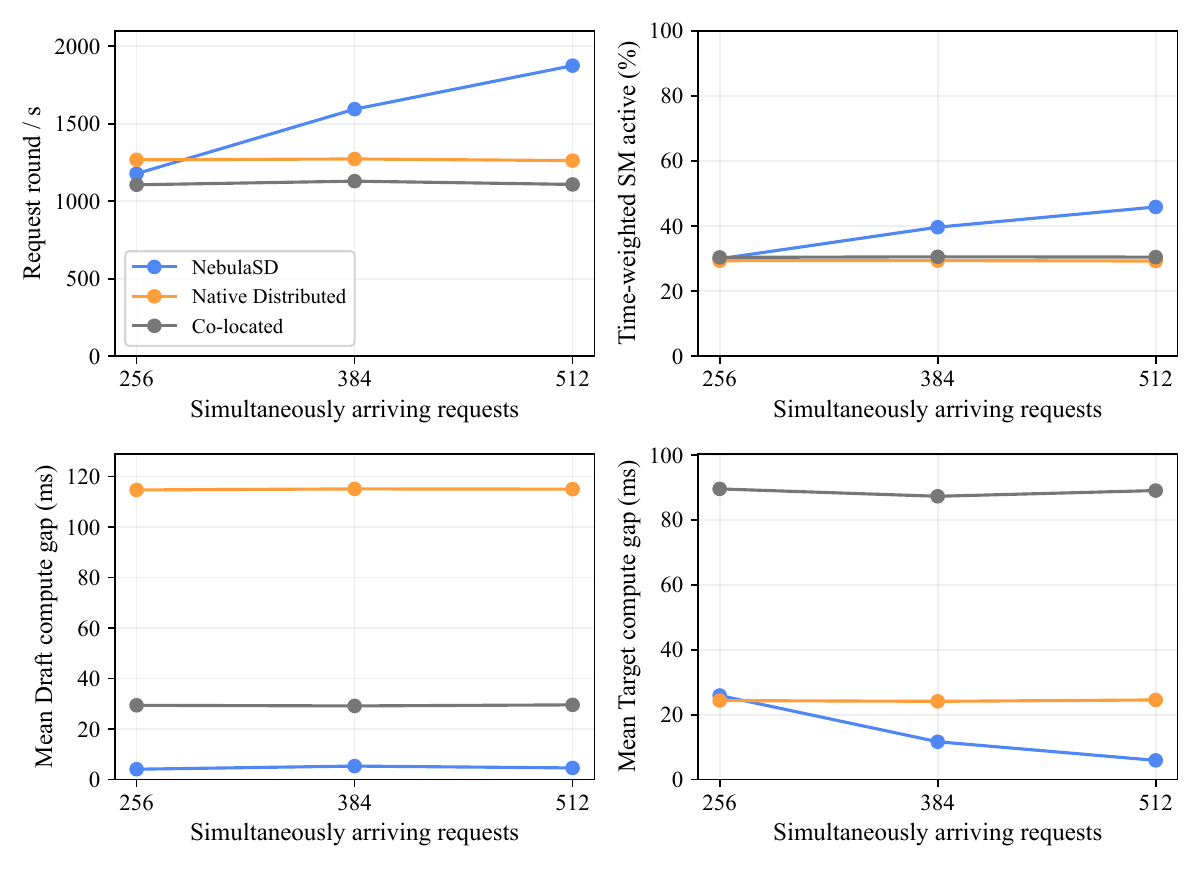}
\caption{Sensitivity to submitted request concurrency. The panels report steady-state request-round processing rate, time-weighted SM activity, mean Draft compute gap, and mean Target compute gap. The same NebulaSD scheduler configuration is used for all request counts without workload-specific retuning.}
\label{fig:request_concurrency}
\end{figure}

Figure~\ref{fig:request_concurrency} shows a clear concurrency-dependent crossover. At 256 submitted requests, Native Distributed slightly outperforms NebulaSD in steady-state request-round processing rate. In this regime, the fixed draft-target pairs already have sufficient local work, and the shared request pools in NebulaSD provide limited additional opportunity for reconstruction to improve batching and worker continuity. As the submitted concurrency increases to 384 and 512 requests, however, NebulaSD's request-round processing rate rises from 1178 to 1594 and 1875 rounds/s, while the fixed-affinity baselines remain close to their saturated rates. NebulaSD consequently exceeds Native Distributed by approximately 25\% at 384 requests and 49\% at 512 requests.

The remaining panels in Fig.~\ref{fig:request_concurrency} explain this crossover from the execution behavior. As the shared request population grows, NebulaSD can select from a larger set of eligible requests and keep both stages more continuously supplied. Its time-weighted SM activity therefore increases with concurrency, while the mean Target compute gap decreases from 25.9 ms at 256 requests to 11.7 and 5.9 ms at 384 and 512 requests, respectively; the Draft compute gap remains consistently small. These trends indicate that the larger request population provides more schedulable flexibility for batch reconstruction rather than merely increasing the amount of outstanding work.

The results also characterize the operating regime targeted by NebulaSD. At relatively low concurrency, fixed worker assignments can already remain sufficiently supplied, leaving limited additional benefit for global pooling. As concurrent demand grows, shared request pools expose more opportunities to reduce worker idle intervals and utilize aggregate stage capacity. Importantly, the same scheduler configuration is used throughout the sweep, so the high-concurrency gains do not rely on workload-specific retuning.

\subsection{Sensitivity to Model Scale}
\label{app:draft_model_scale}

We further evaluate whether the benefit of NebulaSD depends on the Qwen3-0.6B draft model used in the main experiments. We replace it with Qwen3-1.7B while keeping the Qwen3-8B target model, 512 submitted requests, proposal depth of 4, $2\mathrm{D}+2\mathrm{T}$ resource configuration, draft/target maximum batch sizes of 128/32, target service intervals of 160/130 ms, batch-formation slack of 30 ms, and maximum generation length of 128 tokens unchanged. Because changing the draft model can also change the number of useful tokens produced by each speculative round, we use request-round processing rate here to isolate the serving-side processing capacity.

\begin{table}[t]
\centering
\caption{Sensitivity to the draft-model scale under the 512-request workload. Processing rates are completed request rounds/s.}
\label{tab:draft_model_scale}
\small
\setlength{\tabcolsep}{5pt}
\renewcommand{\arraystretch}{1.12}
\begin{tabular}{@{}lrrrrr@{}}
\toprule
\textbf{Draft model} & \textbf{Co-located} & \textbf{Native} & \textbf{NebulaSD} & \textbf{vs. Native} & \textbf{vs. Co-loc.} \\
\midrule
Qwen3-0.6B & $1116.65$ & $1268.98$ & $1883.99$ & $1.48\times$ & $1.69\times$ \\
Qwen3-1.7B & $1130.14$ & $1241.51$ & $1802.22$ & $1.45\times$ & $1.59\times$ \\
\bottomrule
\end{tabular}
\end{table}

Table~\ref{tab:draft_model_scale} shows that increasing the draft-model size slightly changes the absolute round-processing capacity but preserves the benefit of pooled execution. With Qwen3-1.7B, NebulaSD processes 1802 request rounds/s, compared with 1242 rounds/s for Native Distributed and 1130 rounds/s for Co-located execution, corresponding to $1.45\times$ and $1.59\times$ higher processing rates, respectively. These gains are close to the $1.48\times$ and $1.69\times$ advantages observed with Qwen3-0.6B, indicating that the benefit of many-for-many resource pooling is not specific to the 0.6B/8B model pair.

The modest reduction in NebulaSD's round-processing rate is primarily associated with a change in draft-stage service capacity. Replacing Qwen3-0.6B with Qwen3-1.7B increases the mean draft computation time from 89.70 to 93.01 ms and reduces the average Draft batch size from 89.84 to 87.83. In contrast, the target batch size remains 32 and its mean computation time changes only from 28.22 to 28.45 ms. The mean target compute gap consequently increases from 5.83 to 7.09 ms, consistent with the draft stage supplying verification work at a slightly lower rate. Despite this change in the balance between the two stages, NebulaSD continues to maintain a substantial processing-rate advantage over both fixed-affinity baselines.

\section{Simulation Methodology and Extended Scaling Analysis}
\label{app:simulation}

\paragraph{Simulation methodology.} We use a profiling-driven discrete-event simulator to study the compute-side behavior and scalability of NebulaSD beyond the physical deployment evaluated in Section~\ref{sec:eval_four_gpu}. Each request maintains independent draft and target states and progresses through the same alternating stage dependencies as in the real system. The simulator advances a virtual clock between execution and preparation events, applies all state transitions occurring at the same virtual time before scheduling, and separately maintains compute and \textsc{Prepare} opportunities for each worker. Each \textsc{Prepare} opportunity identifies its destination worker, and the simulator applies the worker-triggered batch-reconstruction abstraction described in Section~\ref{sec:system_model} to select its next batch. An ideal migration estimator sets cross-device state-transfer latency to zero, allowing the simulation to isolate the compute-side behavior of the scheduling policy from practical data-movement and runtime overheads.

The simulator does not reproduce the complete physical runtime. In the real implementation, scheduling decisions interact with HostKV publication, state observation, command dispatch, backpressure, migration completion, and readiness retries. The simulator instead applies these state transitions directly in virtual time and assumes zero communication and control-plane latency. It consequently omits PCIe and network transfer time, H2D and D2H contention, inter-process communication (IPC) overhead, kernel-launch jitter, process interference, and physical GPU-memory fragmentation. The simulation should therefore be interpreted as an idealized evaluation of NebulaSD's batching and scheduling behavior rather than a prediction of full-cluster wall-clock performance.

\paragraph{Workload and execution model.}
All requests arrive at $t=0$ in a post-prefill state with an initial anchor token. We set the speculation depth to $d=4$, so that each speculative round generates up to four proposal tokens in the draft stage and verifies the corresponding proposal in the target stage. This depth therefore determines the per-request computation workload of both stages in each round. For the scaling experiments, each request executes 32 speculative rounds. NebulaSD uses a maximum draft batch size of 512, while the maximum target batch size is 24 or 32 depending on the evaluated resource configuration. Compute latency and SM Active are obtained from offline profiling and interpolated across batch sizes under the same speculation depth. The same profiling model is used consistently across all simulated system scales.

\paragraph{Metrics and Co-located baseline.}
We measure all configurations over the common interior virtual-time window $[1,2.5]$ seconds and verify that no request completes within the measurement interval. Let the measurement duration be $\Delta$, the total number of GPUs be $G$, and the number of submitted requests be $R$. For each compute batch $j$, let $\tau_j$ denote its overlap with the measurement window, $s_j$ its profiled SM Active, and $b_j$ its number of requests. The pool-wide time-weighted SM utilization is
\begin{equation}
U
=
\frac{\sum_j \tau_j s_j}{G\Delta},
\label{eq:sim_sm_utilization}
\end{equation}
where idle GPU-time remains in the denominator. The mean request compute-service fraction is
\begin{equation}
S
=
\frac{\sum_j \tau_j b_j}{R\Delta},
\label{eq:sim_request_service}
\end{equation}
where queued requests also remain in the denominator. We additionally report request-round processing rate, defined as the number of request-level target verification rounds completed within the measurement window divided by $\Delta$.

The Co-located baseline uses the same total number of physical GPUs. Each GPU hosts one draft and one target instance and processes a fixed local request cohort through the two stages. A new FIFO cohort is admitted only after the current cohort completes its round, and requests remain assigned to the same GPU. The draft and target batch sizes are both 24 for the T24 configurations and both 32 for the T32 configurations. Consequently, once the number of active requests exceeds the requests that the fixed local cohorts can serve concurrently, only a subset of requests participates in each round while the remaining requests wait behind their assigned GPU. NebulaSD removes this local cohort constraint by reconstructing batches across the global draft and target pools, allowing the available service capacity to be shared across requests more flexibly.

\begin{figure*}[t]
\centering
\begin{subfigure}[t]{0.32\linewidth}
    \centering
    \includegraphics[width=\linewidth]{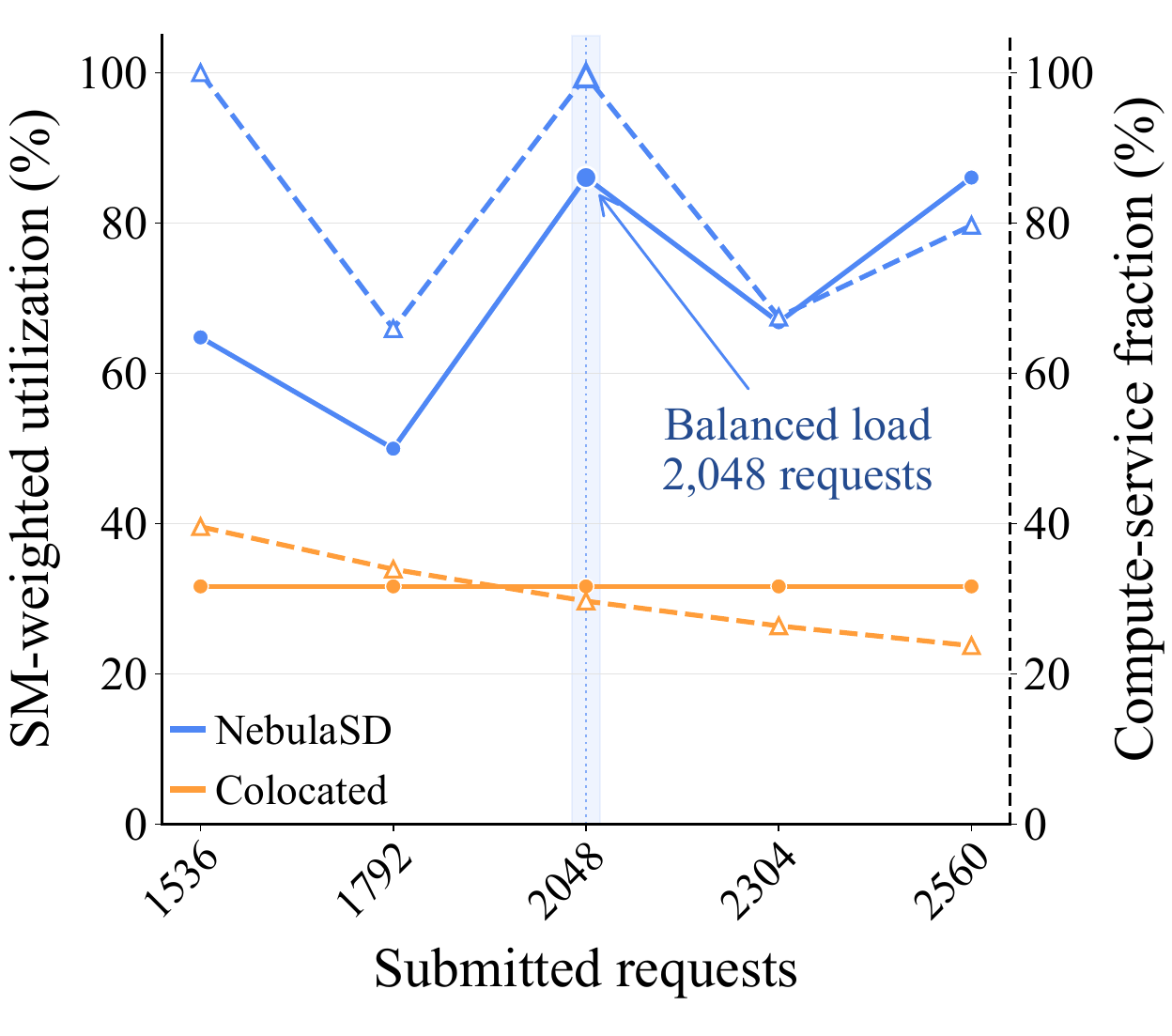}
    \caption{19 GPUs: $3\mathrm{D}+16\mathrm{T}$.}
    \label{fig:sim_app_19gpu}
\end{subfigure}
\hfill
\begin{subfigure}[t]{0.32\linewidth}
    \centering
    \includegraphics[width=\linewidth]{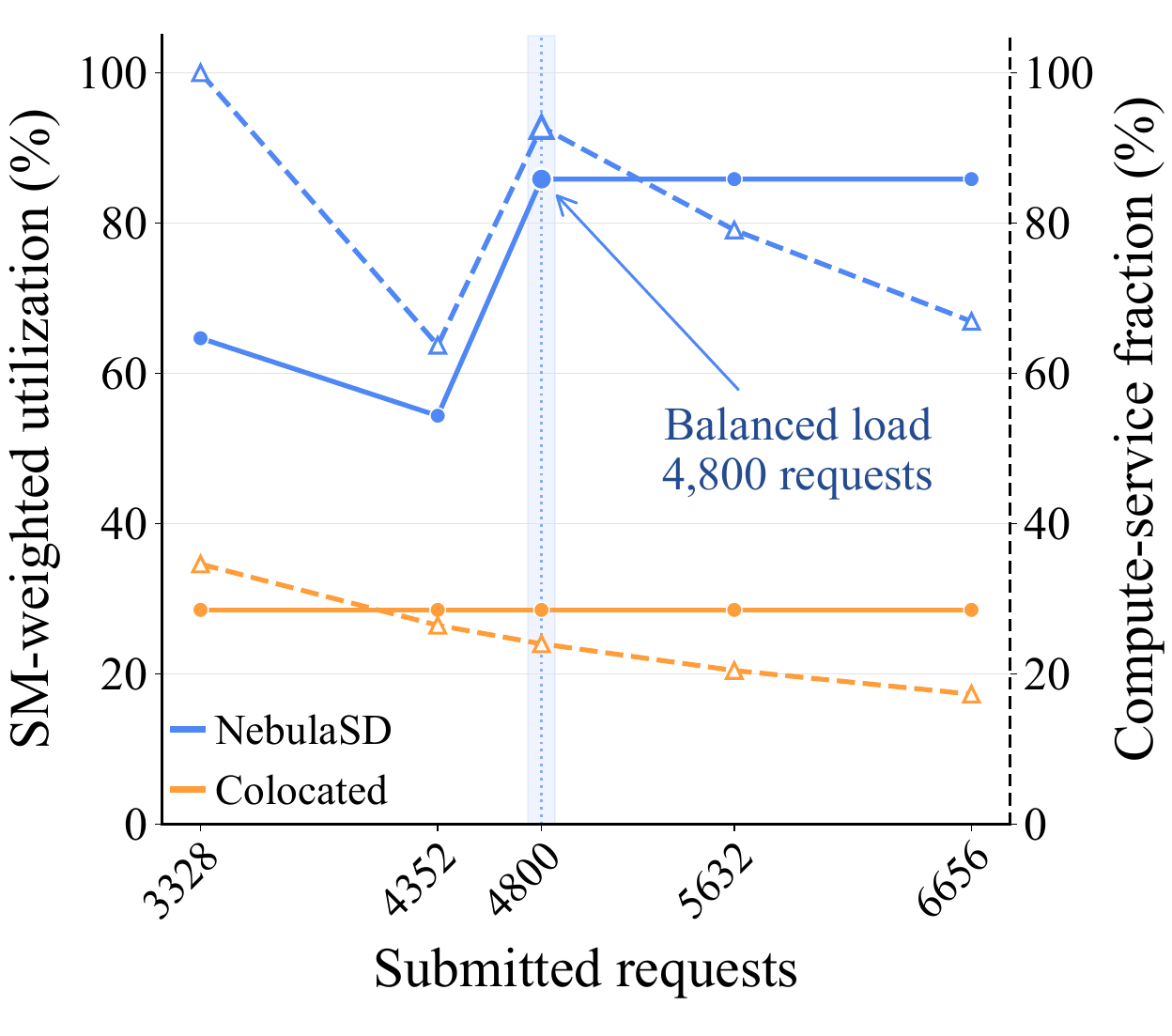}
    \caption{48 GPUs: $8\mathrm{D}+40\mathrm{T}$.}
    \label{fig:sim_app_48gpu}
\end{subfigure}
\hfill
\begin{subfigure}[t]{0.32\linewidth}
    \centering
    \includegraphics[width=\linewidth]{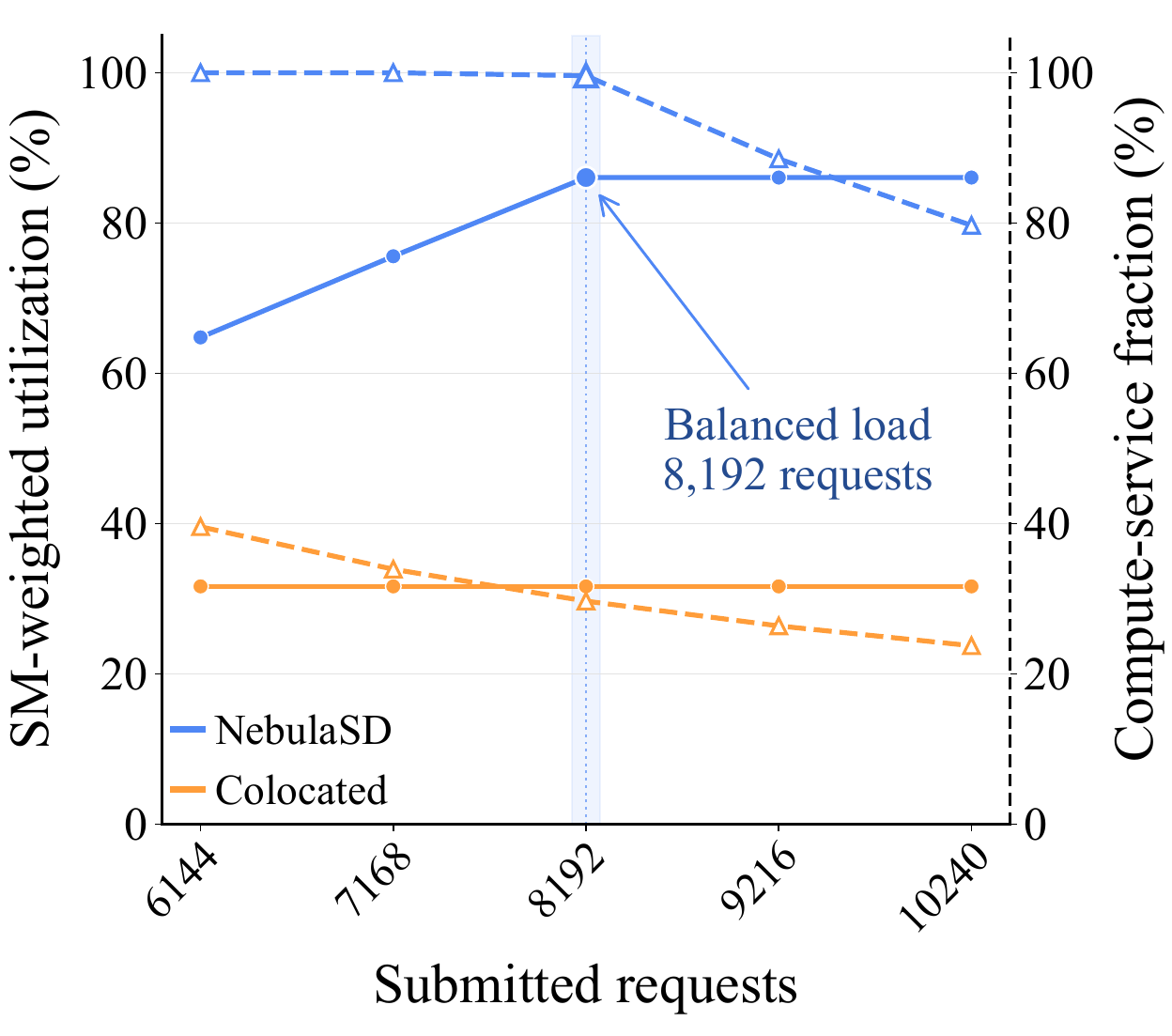}
    \caption{76 GPUs: $12\mathrm{D}+64\mathrm{T}$.}
    \label{fig:sim_app_76gpu}
\end{subfigure}

\vspace{4pt}

\begin{subfigure}[t]{0.32\linewidth}
    \centering
    \includegraphics[width=\linewidth]{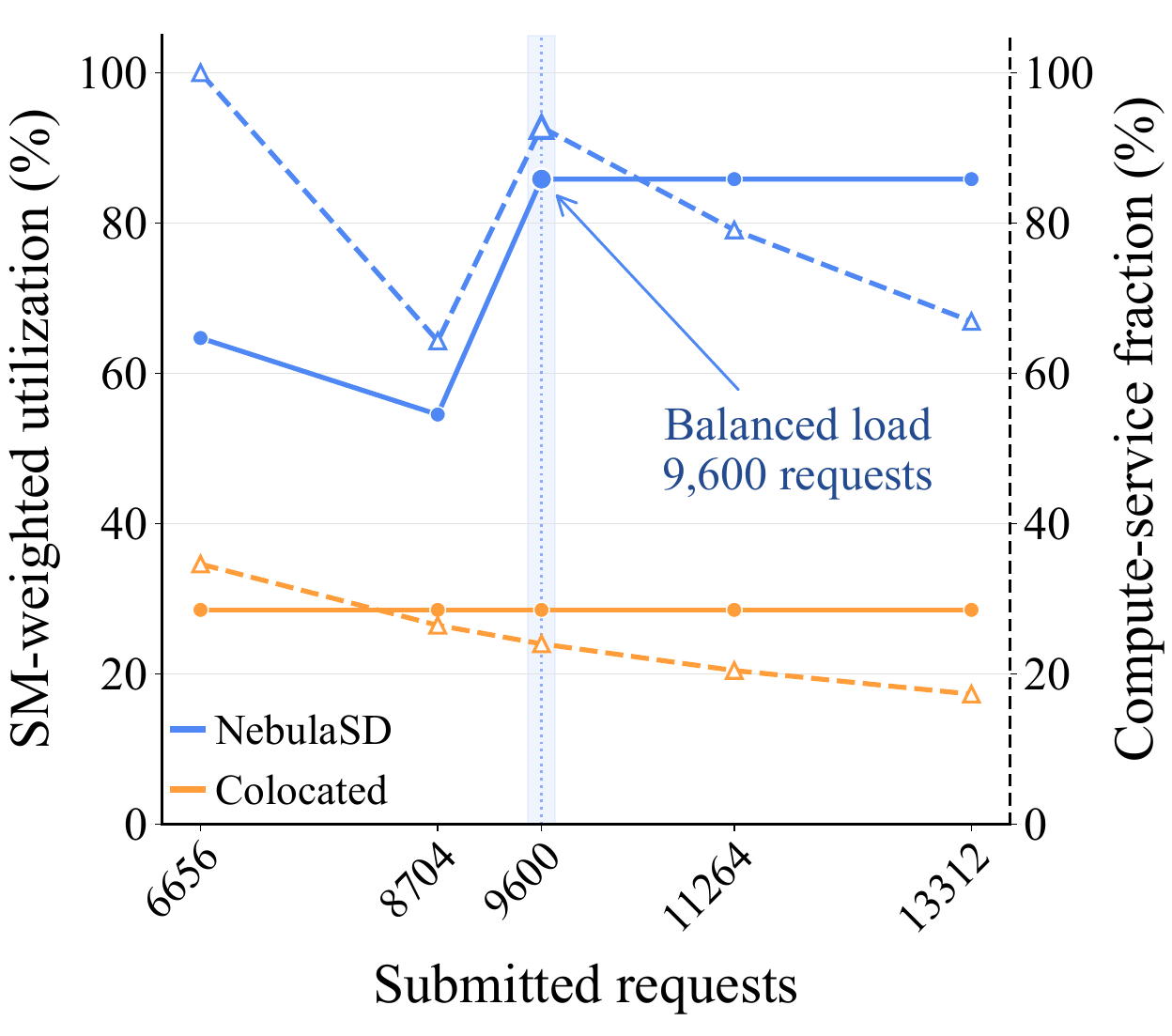}
    \caption{96 GPUs: $16\mathrm{D}+80\mathrm{T}$.}
    \label{fig:sim_app_96gpu}
\end{subfigure}
\hfill
\begin{subfigure}[t]{0.32\linewidth}
    \centering
    \includegraphics[width=\linewidth]{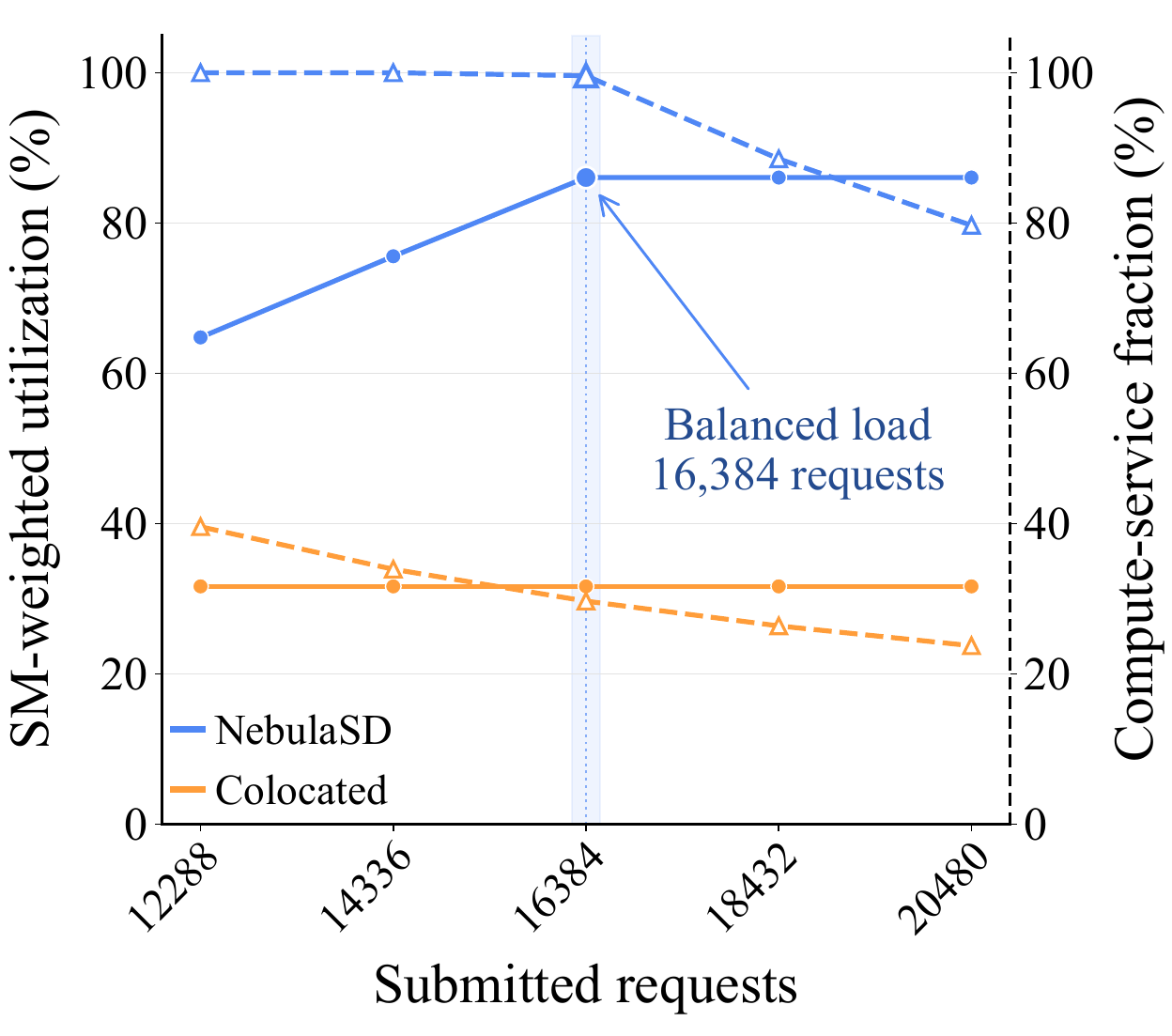}
    \caption{152 GPUs: $24\mathrm{D}+128\mathrm{T}$.}
    \label{fig:sim_app_152gpu}
\end{subfigure}
\hfill
\begin{subfigure}[t]{0.32\linewidth}
    \centering
    \includegraphics[width=\linewidth]{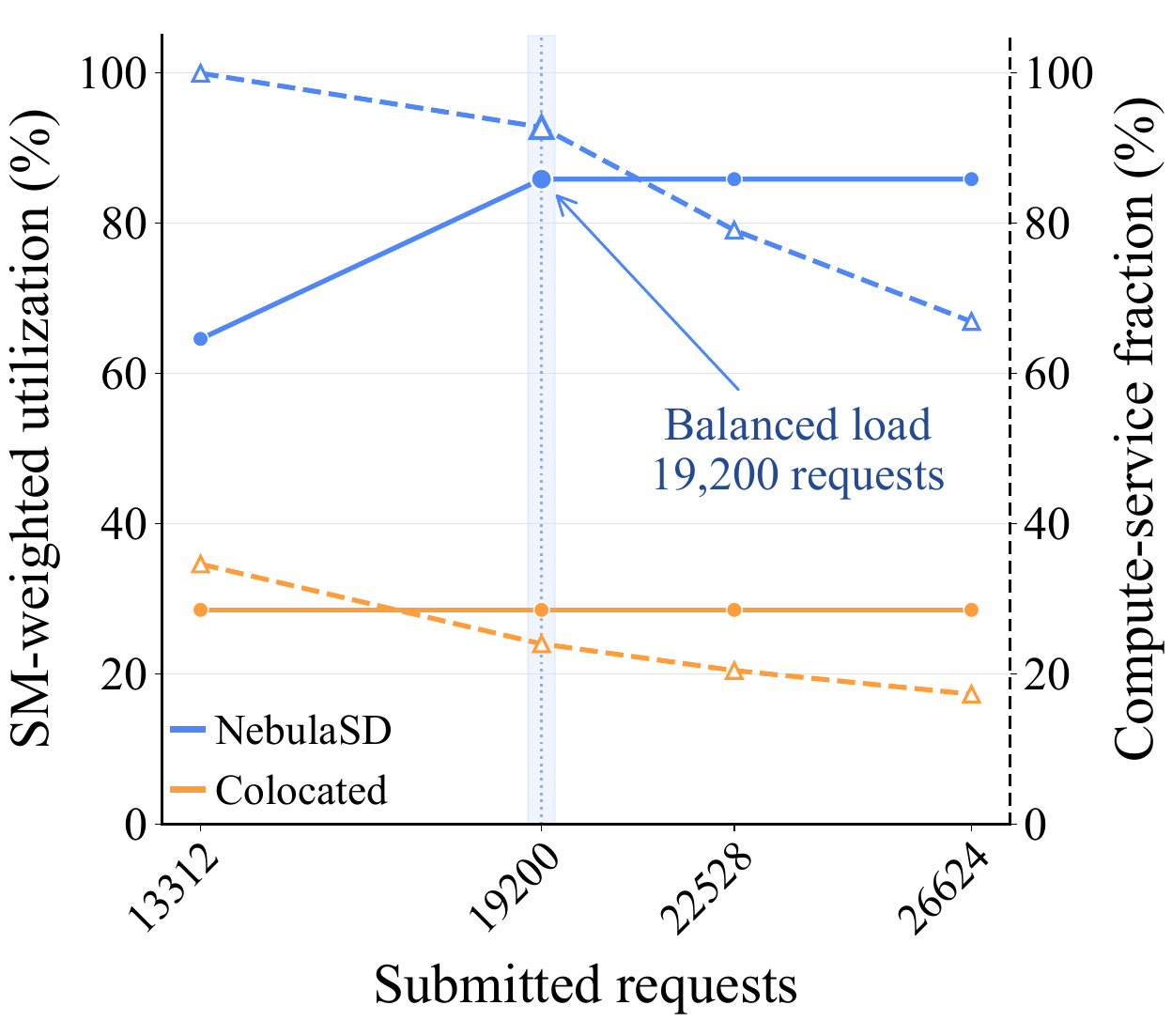}
    \caption{192 GPUs: $32\mathrm{D}+160\mathrm{T}$.}
    \label{fig:sim_app_192gpu}
\end{subfigure}

\caption{Extended simulation results across different system scales and draft-target provisioning ratios. Solid and dashed curves report pool-wide time-weighted SM utilization and mean request compute-service fraction, respectively. Vertical reference lines, where present, indicate the sampled load maximizing $U\times S$ and should be interpreted as a sampled utilization-service trade-off rather than a globally optimal concurrency level.}
\label{fig:sim_extended_scaling}
\end{figure*}

\paragraph{Extended scaling behavior.}
Fig.~\ref{fig:sim_extended_scaling} shows that the utilization-service trade-off observed in the main evaluation persists over substantially larger worker pools. At low request concurrency, individual requests can receive nearly continuous compute service while a large fraction of the GPU pool remains idle. As concurrency increases, larger and more regular stage-specific batches raise hardware utilization without immediately reducing request service quality. Around the favorable sampled operating region, both metrics can remain high even at large scale, as summarized in Table~\ref{tab:sim_extended_points}. For example, the 152-GPU configuration reaches $86.06\%$ time-weighted SM activity and $99.61\%$ mean request service at 16,384 requests while processing 139.3K request rounds/s. The 192-GPU configuration reaches $85.85\%$ time-weighted SM activity and $92.76\%$ request service at 19,200 requests while processing 158.7K rounds/s. For configurations that preserve the same draft-target provisioning ratio and proportionally scale the request population, the idealized homogeneous simulator reproduces the same normalized operating point, causing SM activity and request service to remain unchanged while aggregate round-processing capacity scales proportionally with the worker pool. These results extend the scaling trend reported in Section~\ref{sec:eval_scaling}: enlarging the draft and target pools while preserving an appropriate capacity ratio increases aggregate processing capability without inherently forcing a proportional degradation in per-request service.

\begin{table}[htbp]
\centering
\caption{Representative sampled operating points for the extended scaling configurations. For each configuration, we report the evaluated request count maximizing $U\times S$ among the sampled loads. Rounds/s denotes request-level verification rounds per second.}
\label{tab:sim_extended_points}
\setlength{\tabcolsep}{5pt}
\renewcommand{\arraystretch}{1.08}
\scriptsize
\begin{tabular}{lrrrr}
\toprule
\textbf{Configuration} & \textbf{Requests} & \textbf{SM util.} & \textbf{Service} & \textbf{Rounds/s} \\
\midrule
19 GPU ($3\mathrm{D}+16\mathrm{T}$) & 2,048 & $86.06\%$ & $99.61\%$ & 17,408 \\
48 GPU ($8\mathrm{D}+40\mathrm{T}$) & 4,800 & $85.85\%$ & $92.78\%$ & 39,680 \\
76 GPU ($12\mathrm{D}+64\mathrm{T}$) & 8,192 & $86.06\%$ & $99.61\%$ & 69,632 \\
96 GPU ($16\mathrm{D}+80\mathrm{T}$) & 9,600 & $85.85\%$ & $92.77\%$ & 79,360 \\
152 GPU ($24\mathrm{D}+128\mathrm{T}$) & 16,384 & $86.06\%$ & $99.61\%$ & 139,264 \\
192 GPU ($32\mathrm{D}+160\mathrm{T}$) & 19,200 & $85.85\%$ & $92.76\%$ & 158,720 \\
\bottomrule
\end{tabular}
\end{table}

\paragraph{Interpretation and limitations.}
These simulations characterize the compute-side scalability and utilization-service trade-off of NebulaSD under idealized state movement, rather than the complete performance of a physical large-scale deployment. Communication contention, KV-migration overhead, control-plane cost, and workload variability are intentionally omitted, so the results should be interpreted as the scaling potential of the batch-reconstruction and scheduling policy. Together with the real-system evaluation in Section~\ref{sec:eval_four_gpu}, the two sets of experiments provide complementary views, the physical deployment measures practical runtime performance, while the simulations isolate the behavior of dynamic pooling at larger scales.

\end{document}